\documentclass[letterpaper,12pt]{article}
\usepackage[margin=1in,letterpaper]{geometry}
\usepackage{amsmath}
\usepackage{amssymb}
\usepackage{amsthm}
\usepackage{bbm}
\usepackage{physics}      
\usepackage[normalem]{ulem}
\usepackage{caption}
\usepackage{subcaption}
\usepackage{adjustbox}
\usepackage{comment}
\usepackage[section]{placeins}
\usepackage{mathrsfs}
\usepackage{mathtools}
\usepackage{xcolor}
\usepackage[numbers,sort&compress,square]{natbib}     
\usepackage[final]{hyperref}
\hypersetup{       
	colorlinks=true,
    linkcolor=blue,        
	citecolor=blue,        
	filecolor=magenta,     
	urlcolor=blue,
}

\newcommand{\eq}[1]{\text{Eq.~\ref{eq:#1}}}
\newcommand{\fig}[1]{\text{Fig.~\ref{fig:#1}}}
\newcommand{\sect}[1]{Section~\ref{sec:#1}}
\newcommand{\app}[1]{Appendix~\ref{appendix:#1}}
\newcommand{\create}[1]{\hat{a}^{\dag}(#1)}
\newcommand{\ann}[1]{\hat{a}(#1)}
\newcommand{\fdisp}[1]{\mathcal{D}^{\text{freq}}\left( #1 \right)}
\newcommand{\tdisp}[1]{\mathcal{D}^{\text{time}}\left( #1 \right)}
\newcommand{\ketbfc}{\ket{\varphi_{00}}}
\newcommand{\DOmega}{\Delta\Omega}

\definecolor{mygreen}{rgb}{0.0, 0.6, 0.0}

\begin{document}

\title{GKP-inspired high-dimensional superdense coding with energy-time entanglement}
\author{Kai-Chi Chang$^{1,\dagger,*}$, Arjun Mirani$^{2,4,\parallel,*}$, Murat Can Sarihan$^{1}$, \\
Xiang Cheng$^{1}$, Michelle Harasimowicz$^{3,4}$, Patrick Hayden$^{3,4,\ddagger,*}$, \\
and Chee Wei Wong$^{1,\S}$}
\date{February 16, 2026}
\maketitle

\begin{center}
$^{1}$Fang Lu Mesoscopic Optics and Quantum Electronics Laboratory, Department of Electrical and Computer Engineering, University of California, Los Angeles \\
$^{2}$Department of Applied Physics, Stanford University \\
$^{3}$Department of Physics, Stanford University\\
$^{4}$Leinweber Institute for Theoretical Physics, Stanford University \\[6pt]

$^{*}$These authors contributed equally to this work.\\
$^{\dagger}$\texttt{uclakcchang@ucla.edu}\\
$^{\parallel}$\texttt{asmirani@stanford.edu}\\
$^{\ddagger}$\texttt{phayden@stanford.edu}\\
$^{\S}$\texttt{cheewei.wong@ucla.edu}
\end{center}

\vspace{1em}

\begin{abstract}
Superdense coding, the application of entanglement to boost classical communication capacity, is a cornerstone of quantum communication. In this paper, we propose a high-dimensional superdense coding protocol using energy-time entangled states. These states are biphoton frequency combs, an example of entangled time-frequency Gottesman-Kitaev-Preskill (TFGKP) states or time-frequency grid states. Inspired by GKP codes, our protocol involves discretizing the continuous time and frequency degrees of freedom and encoding information by time-frequency displacements. This approach leverages the inherently large Hilbert space found in quantum frequency combs, with resilience against both temporal and spectral errors. In addition to describing the theoretical structure of the protocol, we propose an experimental implementation using standard telecommunication components, time-resolving single-photon detectors and a frequency beamsplitter. We also analyze the effect of experimental noise and errors on the channel capacity of the protocol. We demonstrate that for realistic experimental parameters, contemporary technologies satisfy the prerequisites for superdense coding with biphoton frequency combs, achieving a transmission rate of approximately 8.91 bits per transmitted photon (equivalent to 481 distinguishable messages with asymptotically vanishing errors). This more than doubles the previously highest transmission rate of 4 bits achieved by the Kwiat-Weinfurter scheme, while also having competitive optical loss. Furthermore, our results beat the rate achievable using a single-photon frequency comb with identical parameters by 4.6 times. Our protocol thus represents an experimentally feasible application of time-frequency grid states to entanglement-assisted communication, contributing to the active fields of continuous-variable and high-dimensional quantum information.
\end{abstract} 

\begingroup
\hypersetup{hidelinks}
\tableofcontents
\endgroup

\section{Introduction}
Superdense coding allows a single qubit to convey a pair of classical bits from sender Alice to receiver Bob by leveraging their shared entanglement resource \citep{Mattle1996,Fang2000,Li2002,Harrow2004,Schaetz2004,Barreiro2008,Williams2017,Hu2018,Graham2015,Chapman2020}. In the simplest version, Alice encodes a pair of classical bits onto her half of a Bell state and subsequently transmits the qubit to Bob. Bob then conducts a comprehensive Bell-state measurement on the dual qubits and deciphers the encoded classical information. Throughout this sequence, Alice transmits only one qubit to Bob, who, in turn, obtains two classical bits of information. This striking application of entanglement has been realized in various physical systems, including atomic \citep{Fang2000,Schaetz2004} and photonic platforms \citep{Li2002,Harrow2004,Barreiro2008,Williams2017, Hu2018}. However, linear optics imposes a maximum success probability of 50\% for Bell-state measurements \citep{Calsamiglia2001}. Both hyperentanglement \citep{Barreiro2008,Williams2017,Hu2018}, and high-dimensional \citep{Hu2018} quantum states have been utilized to overcome this hurdle in order to achieve higher communication rates.

However, in recent hyperentanglement-based superdense coding schemes \citep{Barreiro2008,Williams2017,Hu2018}, the time and frequency degrees of freedom have remained unexplored. Utilizing time and frequency offers numerous benefits. As continuous variables, these degrees of freedom are well-suited for implementing high-dimensional quantum information processing \citep{Xie2015,Jaramillo-Villegas2017,Ikuta2019,Imany2019,Reimer2019,Maltese2020,Chang2021,Chang2025a,Chang2025b,Chang2023,Chang2023b}. Furthermore, information transmitted through time and frequency exhibits greater error resilience, as optical components typically remain unaffected by minor temporal and spectral variations.

\noindent
\textbf{Our contribution}:

\noindent
In this paper, we propose an experimentally viable superdense coding protocol using the time and frequency degrees of freedom of biphoton frequency combs, which are energy-time entangled states. In addition to describing the theory and experimental setup in detail, we analyze the information transmission rate of the protocol for realistic experimental parameters. Our estimated rate of 8.91 bits per transmitted photon more than doubles the state-of-the-art rate of 4 bits achieved by Kwiat-Weinfurter scheme \citep{Kwiat1998,Wei2007,Wang2019}. 

Our protocol uses biphoton frequency combs, which are examples of entangled time-frequency Gottesman-Kitaev-Preskill (TFGKP) states or time-frequency grid states \citep{Fabre2020, Descamps2024}. The mathematical structure of our protocol is inspired by the continuous-variable GKP code in quantum error correction, which encodes a qudit into a quantum harmonic oscillator via discrete phase-space displacements, and prior theoretical work on continuous-variable superdense coding based on squeezed vacuum electromagnetic states \citep{Ban1999, BraunsteinKimble2000}. The biphoton frequency combs used in our protocol have EPR-like entanglement in the continuous time and frequency variables, forming the entanglement resource shared between sender Alice and receiver Bob. Alice encodes classical information by applying time and frequency displacements to the photon in her possession. She then transmits the photon to Bob. The decoding stage involves Bob inferring these displacements, by first sending both photons through a frequency beamsplitter (a continuous-variable version of a symmetric CNOT gate acting on the frequency DoF), followed by measuring one of the photons in the frequency basis and the other in the temporal basis. It is worth mentioning that our protocol is \textit{not} simply textbook superdense coding with generalized logical Pauli operators applied to logical Bell states of independently-encoded GKP qudits. Compared to the latter, the algebraic structure of our biphoton comb states allows for a higher encoding capacity for the same experimental parameters, as we discuss later on.

In addition to describing the theory of how our protocol works, we propose a detailed experimental implementation. The biphoton comb that forms the shared entangled resource between Alice and Bob can be implemented through a nonlinear optical process assisted by a cavity \citep{Xie2015,Jaramillo-Villegas2017,Ikuta2019,Imany2019,Reimer2019,Chang2021,Chang2025a, Chang2023}. Its EPR-like wavefunction has support on a discrete series of spectral peaks, whose control and manipulation has been shown using a series of electro-optic modulators and pulse shapers \citep{Imany2018,Lu2018,Seshadri2022}. Hence the encoding stage is implemented by electro-optic modulators (EOMs), tunable dispersion modules and telecom fibers. The decoding stage involves applying a frequency beamsplitter operation, which can be implemented near-deterministically with linear optics, followed by frequency and time measurements using tunable frequency filters and single-photon detectors.

Moreover, different from earlier hyperentanglement approaches \citep{Barreiro2008,Williams2017,Hu2018}, the superdense coding proposed here remains resilient even in the presence of finite detector resolutions and various temporal and spectral errors. Although the theory section describes our protocol in idealized terms, we subsequently assess the possible errors present in this protocol for realistic experimental parameters, and show that they can be overcome using a layer of classical error correction. We estimate the resulting transmission capacity of our scheme to be 8.91 bits per transmitted photon, equivalent to 481 distinguishable messages when considering asymptotically vanishing errors. This exceeds the rate achieved by a single-photon frequency comb with the same parameters by 4.6 times, as well as the highest previously achieved superdense coding capacity of the Kwiat-Weinfurter scheme by 2.2 times \citep{Kwiat1998,Wei2007,Wang2019}.

\noindent
\textbf{Outline of the paper:}

\noindent
The rest of the paper is structured as follows. \sect{tfgkp_review} summarizes the essential theory background needed for our protocol. After reviewing the quantum mechanics of time and frequency as canonically conjugate variables, we describe the key properties of single-photon frequency combs and time-frequency GKP coding in an easily visualizable setting, providing motivation and intuition relevant to the entangled two-photon case of our protocol.

Next, in \sect{sdc_protocol}, we describe an idealized version of our superdense coding protocol based on biphoton frequency combs, whose correlation structure is similar in some ways to the single-photon TFGKP states discussed in \sect{tfgkp_review}, yet different in crucial ways that we highlight. After introducing biphoton comb states in \sect{bfc_intro}, we describe the encoding stage of the superdense coding protocol in \sect{sdc_encoding} and the decoding stage in \sect{sdc_decoding}. Since this section describes the protocol in terms of idealized, hence unphysical, comb states for analytical clarity, we describe in \sect{bfc_envelopes} how to modify the formalism to account for physical effects, setting the stage for the subsequent experimental sections.

In \sect{exp_design}, we describe the proposed experimental implementation of our protocol. The generation of the initial biphoton frequency comb state is described in \sect{exp_generation}, the implementation of the frequency beam splitter operation in \sect{exp_fbs}, and the various steps of encoding and decoding in \sect{exp_encdec}.

In \sect{channel_capacity}, we estimate the channel capacity (transmission rate) of our superdense coding protocol for the experimental design described in the previous section. In \sect{exp_params}, we specify the sources and numerical values of experimental errors/noise for current state-of-the-art hardware. In \sect{capacity_calc}, we explain how applying a layer of classical error correction allows for an asymptotically vanishing error rate, and compute the resulting channel capacity of our protocol. Then, in \sect{comparisons}, we make three natural comparisons of our protocol's transmission capacity to that of other ways of transmitting classical information using quantum states: single-photon frequency combs; the qudit version of textbook superdense coding implemented with logical Bell states of TFGKP qudits; and prior schemes from the superdense coding literature. As mentioned above, our protocol improves upon all of these methods.

Finally, we conclude in \sect{conclusion}. \app{app_a} provides an explicit calculation of the Fourier transform to express the biphoton frequency comb in the temporal basis, while \app{app_b} provides some alternative experimental configurations, including a superdense coding protocol based on frequency-polarization hyperentanglement.  

\section{Time-frequency Gottesman-Kitaev-Preskill (TFGKP) states}\label{sec:tfgkp_review}
In this section, we summarize the conceptual ingredients needed for our TFGKP-inspired superdense coding protocol. 

\subsection{Review of time and frequency as quantum canonical conjugates}\label{sec:tf_review}
Before introducing time-frequency GKP (henceforth TFGKP) states, we briefly review the key properties of frequency and time as canonically conjugate continuous-spectrum quantum observables. This subsection primarily draws upon \citep{Fabre2022a}, where more details can be found.

The frequency operator $\hat{\omega}$ is defined in terms of creation and annihilation operators, $\create{\omega}$ and $\ann{\omega}$, as
\begin{equation}
    \hat{\omega} = \int_{-\infty}^{\infty} \omega \create{\omega} \ann{\omega}
\end{equation}
While the integral formally runs over all of $\mathbb{R}$, every physical state will have a finite bandwidth in the positive frequency regime.

Similarly, the time degree of freedom, which refers to the time of arrival at a specified detector, is defined in terms of creation and annihilation operators $\create{t}$ and $\ann{t}$ as follows:
\begin{equation}
    \hat{t} = \int_{-\infty}^{\infty} t \create{t} \ann{t}
\end{equation}

The creation and annihilation operators for frequency and time are related by the Fourier transform:
\begin{equation}
    \create{\omega} = \frac{1}{\sqrt{2\pi}}\int dt \; e^{-i\omega t}\create{t}
\end{equation}

As shown in \citep{Fabre2022a}, when restricted to the subspace of Fock space in which each mode is occupied by at most one photon, the operators $\hat{\omega}$ and $\hat{t}$ satisfy the canonical commutation relation:
\begin{equation}
    [\hat{\omega}, \hat{t}] = i\mathbb{I}
\end{equation}
Thus $\hat{\omega}$ and $\hat{t}$ are canonical conjugates that behave analogously to position and momentum, or the quadratures of the electromagnetic field. Throughout this paper, we will work in the single-photon-per-mode subspace where the canonical commutation relation holds. We will use the following shorthand ket notation for frequency and time eigenstates respectively:
\begin{align}
    \ket{\omega}^{\text{freq}} &= \create{\omega}\ket{\text{v.s.}}\\
    \ket{t}^{\text{time}}  &= \create{t}\ket{\text{v.s.}}
\end{align}
where $\ket{\text{v.s.}}$ is the vacuum state. Our Fourier transform conventions imply that the temporal-basis wavefunction of a frequency eigenstate is $\braket{t}{\omega} = \frac{1}{\sqrt{2\pi}}e^{i\omega t}$.

\subsection{Single-photon TFGKP states and displacement operators}\label{sec:single_photon}
Our superdense coding protocol utilizes biphoton frequency combs, which will be introduced in \sect{sdc_protocol}. However, in this section, we first introduce \textit{single-photon} time-frequency comb states. As observed in \citep{Fabre2020}, these are instances of time-frequency GKP (TFGKP) states, analogous to the usual GKP encoding of a logical qudit into grid states of a quantum harmonic oscillator \citep{Gottesman2001}. We review the key properties of these states to illustrate GKP-based encoding via displacement operators in an easily visualizable setting. 

Before diving in, we note that, for our purposes, the GKP-coding interpretation of time-frequency combs is mainly for motivation and intuition. The entangled biphoton combs used in our protocol, which are EPR-like entangled TFGKP states, are \textit{not} simply logical Bell states of independently encoded TFGKP qudits. The differences between our biphoton combs and conventional logical GKP states play an important role in our protocol and will be explained in the next section.

Now we review single-photon time-frequency comb states. The idealized wavefunction of a single photon comb comprises an infinite series of periodically spaced peaks, in both the frequency and temporal bases. In other words, the state is a Dirac comb, which in the frequency basis can be expressed as:
\begin{align}\label{eq:freq_sp_comb}
\ket{\psi_0}_f &= \sum_{m=-\infty}^{\infty} \int_{-\infty}^{\infty} d\Omega \, \delta(\Omega - m\Delta\Omega) \create{\omega_0 + \Omega} \ket{\text{v.s.}} \nonumber \\
&= \sum_{m=-\infty}^{\infty} \ket{\omega_0 + m\Delta\Omega},
\end{align}
where $\ket{\text{v.s.}}$ is the vacuum state, $\Delta\Omega$ is the frequency-basis periodicity of the comb, and $\omega_0$ is some central frequency. Since this state is not normalizable, in subsequent manipulations we will omit overall constants. 

The temporal basis state, given by the Fourier transform of \eq{freq_sp_comb}, has the same functional form of a Dirac comb, but with periodicity $\Delta T = 2\pi/\Delta\Omega$:
\begin{align}\label{eq:time_sp_comb}
\ket{\psi_0}_t &= \sum_{m=-\infty}^{\infty} \int_{-\infty}^{\infty} dt \, \delta(t - m\Delta T) \create{t_0 + t} \ket{\text{v.s.}} \nonumber \\
&= \sum_{m=-\infty}^{\infty} \ket{t_0+m\Delta T}
\end{align}

The Dirac comb state of \eq{freq_sp_comb} and \eq{time_sp_comb}, can be equivalently expressed as a grid-like Wigner function on the chronocyclic (time-frequency) phase space, with nontrivial support on a discrete infinite lattice of points \citep{Fabre2020, Descamps2023}. While occasionally referring to the phase space picture for intuition, we will use the wavefunction formalism for all our calculations. 

One can define displacement operators that apply frequency and time translations:
\begin{align}
    \fdisp{\Delta\omega} &\coloneq \int_{-\infty}^{\infty} d\omega \, \create{\omega+\Delta\omega} \ann{\omega}\\
    \tdisp{\Delta t} &\coloneq \int_{-\infty}^{\infty} dt \, \create{t+\Delta t} \ann{t}
\end{align}
In phase space, these can be viewed as implementing translations of the lattice, along the frequency and time axes respectively. 

These displacement operators form a representation of the continuous Heisenberg-Weyl group, and satisfy the corresponding commutation relation \citep{Fabre2022a}:
\begin{equation}
    \fdisp{\Delta\omega}\tdisp{\Delta t} = e^{i(\Delta\omega)(\Delta t)} \; \tdisp{\Delta t} \fdisp{\Delta\omega}
\end{equation}

The TFGKP encoding \citep{Fabre2020}, analogous to the original GKP encoding for quantum harmonic oscillators \citep{Gottesman2001}, encodes a logical qudit into a subspace of Dirac comb states by applying a discrete set of displacement operators to some initial comb. Concretely, a set of orthogonal comb states with the same period can be interpreted as logical $Z$-basis states of a $d$-dimensional qudit as follows. First, divide the period $\Delta\Omega$ of the comb in the frequency basis into discrete bins of equal size (alternatively, one can start with the time basis and follow the subsequent steps analogously). Let the bin size be $\Delta\Omega/d$, where $d$ is the qudit dimension. This effectively discretizes the continuous frequency variable into $d$ bins per period. Now, define a frequency displacement by $\Delta\Omega/d$ to be the logical $X$ operator:
\begin{equation}\label{eq:logical_X}
    X\hat{a}_{H}^{\dagger}(\omega)\ket{\text{v.s.}} \coloneq \hat{a}_{H}^{\dagger}\left(\omega + \frac{\Delta\Omega}{d}\right)\ket{\text{v.s.}}
\end{equation}

\begin{figure}
    \centering
    \includegraphics[width=0.75\columnwidth]{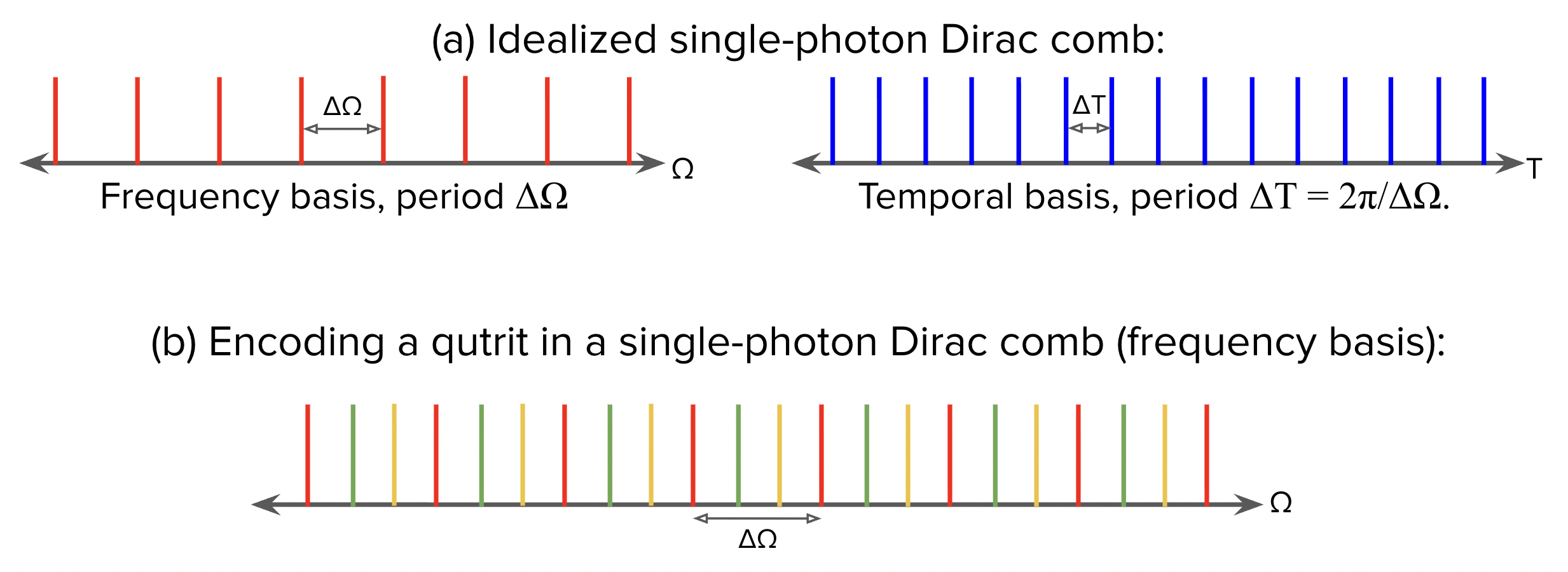}
    \caption{(a) Schematic illustration of a single-photon Dirac comb state in the frequency basis (\eq{freq_sp_comb}) and the same state in the temporal basis (\eq{time_sp_comb}). Each spike represents a Dirac delta function. (b) GKP encoding of the three basis states of a logical qutrit in the frequency basis of a single-photon Dirac comb, by discrete frequency shifts of $\frac{\Delta\Omega}{3}$ that translate the entire comb. The red comb represents the logical $\ket{\overline{0}}$ state, the green comb represents $\ket{\overline{1}}$ and the yellow comb represents $\ket{\overline{2}}$.}
    \label{fig:gkp_qutrit}
\end{figure}

Now, fix some initial comb state, say $\ket{\psi_0}_f$ from \eq{freq_sp_comb}, to be the $\ket{\overline{0}}$ state of the qudit (the overline denotes logical qudit states). The discrete orbit of orthogonal comb states generated by the action of the operators $\{X^j \; | \; j \in \{0,\dots,d-1\}\}$ on $\ket{\psi_0}_f$ correspond, in the language of GKP codes, to the basis states $\{\ket{\overline{j}} \; | \; j \in \{0,\dots,d-1\}\}$:
\begin{equation}\label{eq:z_eigstates}
    \ket{\overline{j}} \coloneq X^j\ket{\psi_0} = \sum_{m=-\infty}^{\infty} \ket{\omega_0 + \left(m + \frac{j}{d}\right)\Delta\Omega}
\end{equation}
which is the natural generalization of the usual Pauli $X$ operator to higher-dimensional qudits. The set of states described by \eq{z_eigstates} is schematically illustrated for the case of a qutrit ($d$=3) in \fig{gkp_qutrit}. 

One can check that for a logical $Z$ operator defined to be a time displacement by $2\pi/\Delta\Omega$,
\begin{equation}\label{eq:logical_Z}
    Z\hat{a}_{H}^{\dagger}(t)\ket{\text{v.s.}} \coloneq \hat{a}_{H}^{\dagger}\left(t + \frac{2\pi}{\Delta\Omega}\right)\ket{\text{v.s.}}
\end{equation}
it acts on the logical basis states defined in \eq{z_eigstates} as the appropriate generalization of the Pauli $Z$ operator,
\begin{equation}
    Z \ket{\overline{j}} = (e^{-\frac{2\pi i}{d}})^j \; \ket{\overline{j}}
\end{equation}
These $X$ and $Z$ operators satisfy the discrete Heisenberg-Weyl commutation relation $ZX = e^{-2\pi i/d}XZ$ on the TFGKP subspace, as well as $X^d = Z^d = \mathbb{I}$ on this subspace, thereby providing a representation of the generators of the $d$-dimensional finite Heisenberg-Weyl group. 

Before moving on to our protocol, we highlight a physically relevant aspect of the TFGKP encoding defined by \eq{z_eigstates}. The number of orthogonal states that can be encoded in such comb states, and hence the dimension of the logical qudit, is limited by two quantities: the finite period and finite bin width, where the latter is ultimately limited by the finite resolution of physical operations such frequency shifts and measurements. In the next section, we will see how biphoton frequency combs, while in some respects analogous to the single-photon combs described above, circumvent the limitation of the finite period in the frequency basis and allow more orthogonal states to be encoded.

\section{Biphoton frequency comb-based superdense coding protocol}\label{sec:sdc_protocol}

\begin{figure}
    \centering
    \includegraphics[width=0.65\columnwidth]{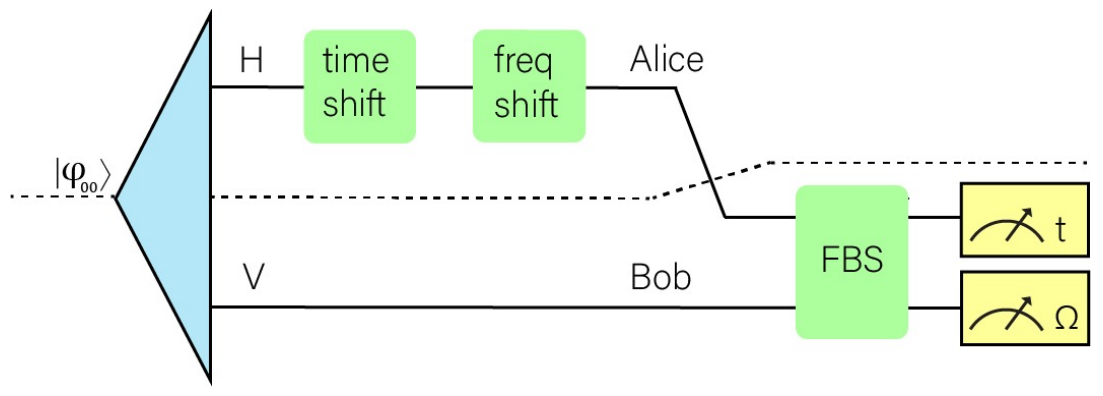}
    \caption{Schematic representation of the superdense coding protocol based on FBS decoder. The final measurements are of the arrival-time of one photon and the frequency of the other.}
    \label{fig:protocol_schematic}
\end{figure}

Superdense coding comprises two stages: encoding performed by the sender, and decoding performed by the receiver. In our protocol, the sender and receiver each initially possess one photon of the entangled pair described by \eq{bfc_dirac}. The encoding process, inspired by GKP coding, involves applying temporal and spectral shifts to the sender's photon. The decoding process utilizes a frequency beamsplitter (FBS), with subsequent measurements of one photon in the frequency basis and the other in the temporal basis, performed by the receiver. These measurements enable the determination of the initial state's temporal and spectral shifts, distinguishing between various encoded states, interpreted as distinct classical messages. Our protocol is schematically illustrated in Fig.~\ref{fig:protocol_schematic}.

At this point, it is worth pointing out that our protocol is \textit{not} simply textbook superdense coding implemented with generalized Pauli operators acting on logical ``Bell pair" of independently encoded TFGKP qudits of the form $1/\sqrt{d}(\ket{\overline{0}}\ket{\overline{0}}+\dots + \ket{\overline{d-1}}\ket{\overline{d-1}})$. Our physically motivated biphoton frequency combs, while having EPR-like correlations, have a different algebraic structure from such logical Bell states, such that the former provide a higher channel capacity for the same experimental parameters, as we will see shortly.

Below, as in \sect{single_photon}, we initially describe our protocol in terms of idealized, non-normalizable states for analytical clarity. After describing the encoding and decoding steps, we explain how to modify our calculations to account for physical effects, setting the stage for the next section on experimental feasibility.

\subsection{Initial state: biphoton frequency comb}\label{sec:bfc_intro}

The initial state for our protocol, a biphoton frequency comb that can be considered an entangled TFGKP state, is a frequency- (and time-) entangled state of two photons with orthogonal polarizations. It can be expressed in the frequency basis in a form analogous to the Dirac comb: 
\begin{align}\label{eq:bfc_dirac}
\ket{\varphi_{00}} &= \sum_{m=-\infty}^{\infty} \int_{-\infty}^{\infty} d\Omega \, \delta(\Omega - m\Delta\Omega) \hat{a}_{H}^{\dagger}\left(\frac{\omega_p}{2} + \Omega\right) \hat{a}_{V}^{\dagger}\left(\frac{\omega_p}{2} - \Omega\right) \ket{\text{v.s.}} \nonumber \\
&= \sum_{m=-\infty}^{\infty} \left|\frac{\omega_p}{2} + m\Delta\Omega\right\rangle_H^{\text{freq}} \left|\frac{\omega_p}{2} - m\Delta\Omega\right\rangle_V^{\text{freq}} 
\end{align}
$\hat{a}_{H}^{\dagger}$ and $\hat{a}_{V}^{\dagger}$ are the creation operators for horizontally and vertically polarized photons respectively. The subscripts on the kets label the polarizations of the corresponding photons.

The state $\ketbfc$ is an entangled two-photon analog of the Dirac comb of \eq{freq_sp_comb} with period $\Delta\Omega$. It is an EPR-like state analogous to a discretized infinitely squeezed two-mode squeezed state. Such a state can be created in a type-II spontaneous parametric down-conversion (SPDC) configuration, which we discuss further in \sect{exp_design} on experimental design.

In the single photon case of \sect{single_photon}, the Dirac comb had the same functional form in both the frequency and time bases, with the only difference being the period. However, for the entangled comb of \eq{bfc_dirac}, the frequency-basis state does \textit{not} have the same functional form in the temporal basis. This can be checked by explicitly performing the Fourier transform (provided in \app{app_a}), which yields the following state:
\begin{equation}\label{eq:bfc_time}
    \sum_m \int dt \int d\tau\,
  e^{-i\frac{\omega_p}{2}(2t-\tau)}
   \delta(\tau-m\Delta T)
   a_H^\dagger(t)
   a_V^\dagger(t-\tau) \ket{\mathrm{v.s.}}
\end{equation}
where $\Delta T  = 2\pi/\Delta\Omega$.
The asymmetry between the functional forms of the wavefunction in the frequency and temporal bases makes manifest how it behaves under frequency and time shifts that act on only one of the photons. In particular, one can check that the entangled biphoton comb is an eigenstate of single-photon time shifts by $\Delta T$, but it is not an eigenstate of any single-photon (non-zero) frequency shift, unlike the single-photon comb of \eq{freq_sp_comb} and unlike a logical TFGKP Bell state of the form $1/\sqrt{d}(\ket{\overline{0}}\ket{\overline{0}}+\dots + \ket{\overline{d-1}}\ket{\overline{d-1}})$. This asymmetry allows for more orthogonal states to be encoded, in a GKP-like manner, in frequency than in time. We will return to this point in \sect{sdc_encoding} below.

\subsection{Encoding}\label{sec:sdc_encoding}
As mentioned above, the encoding process involves applying single-photon frequency and temporal shifts to the initial biphoton comb state $\ketbfc$ from \eq{bfc_dirac}. The number of distinguishable shifts determines the number of messages that can be encoded. If up to $c$ distinguishable frequency shifts and up to $d$ distinguishable temporal shifts can be applied to $\ketbfc$, then Alice can encode any one of $cd$ messages to transmit to Bob, corresponding to $\log(c) + \log(d)$ bits of classical information. The upper limits $c$ and $d$ on the number of distinguishable shifts ultimately depends on physical considerations, as we discuss in the sections below.

First consider the time domain. The state $\ketbfc$ is an eigenstate of temporal shift operators $\tdisp{s}$ for all $s$ satisfying $s = 0$ mod $2\pi/\Delta\Omega$. Consequently, following discretization of the time variable, the upper limit on the number of distinguishable temporal shifts that can be applied to $\ketbfc$ is set by the number of time bins that fit within the time-domain period $2\pi/\Delta\Omega$. The time bin width, in turn, is set by the finite resolution of temporal shifts and measurements, which we analyze explicitly in \sect{exp_design}.

For now, let the time bin width be $2\pi/(d\Delta\Omega)$, which is $1/d$ times the time-domain period for some $d \in \mathbb{Z}$. This corresponds to $d$ distinct messages. To encode the $j^{\text{th}}$ message, where $j \in \{0,\dots,d-1\}$, Alice applies the displacement operator 
\begin{equation}
    \left(\tdisp{\frac{2\pi}{d\Delta\Omega}}\right)^j = \tdisp{\frac{j}{d}\frac{2\pi}{\Delta\Omega}}
\end{equation}
to her photon $H$. In the frequency basis, this applies a phase-shift,
\begin{equation}
\tdisp{\frac{j}{d}\frac{2\pi}{\Delta\Omega}}\hat{a}_{H}^{\dagger}(\omega)\ket{\text{v.s.}} \propto e^{-i\frac{j}{d}\frac{2\pi\omega}{\Delta\Omega}} \hat{a}_{H}^{\dagger}(\omega)\ket{\text{v.s.}},
\end{equation}
where the sign convention for the exponent follows from $\langle t|\omega\rangle = \frac{1}{\sqrt{2\pi}}e^{i\omega t}$. Therefore the action of $\tdisp{\frac{j}{d}\frac{2\pi}{\Delta\Omega}}$ on the initial state $\ketbfc$ is the following (as usual, up to overall phases and constants):
\begin{align}
\ket{\varphi_{0j}} & \coloneq \tdisp{\frac{j}{d}\frac{2\pi}{\Delta\Omega}}\ketbfc \\
& \propto \sum_{m=-\infty}^{\infty} e^{-2\pi imj/d} \left|\frac{\omega_p}{2} + m\Delta\Omega\right\rangle_H^{\text{freq}}  \left|\frac{\omega_p}{2} - m\Delta\Omega\right\rangle_V^{\text{freq}} 
\end{align}

Next, consider the frequency domain. Since $\ketbfc$ is not an eigenstate of single-photon shifts $\fdisp{s}$ for any $s\neq0$, and furthermore $\fdisp{s}\ketbfc$ is orthogonal to $\ketbfc$ for $s\neq 0$, arbitrary frequency shifts are distinguishable in principle. In practice, the range of distinguishable frequency shifts is limited by the available bandwidth, as well as the discretized frequency bin width. The maximum number of distinguishable shifts is then set by the number of frequency bins that fit within the bandwidth.

\begin{figure}
    \centering
    \includegraphics[width=0.65\columnwidth]{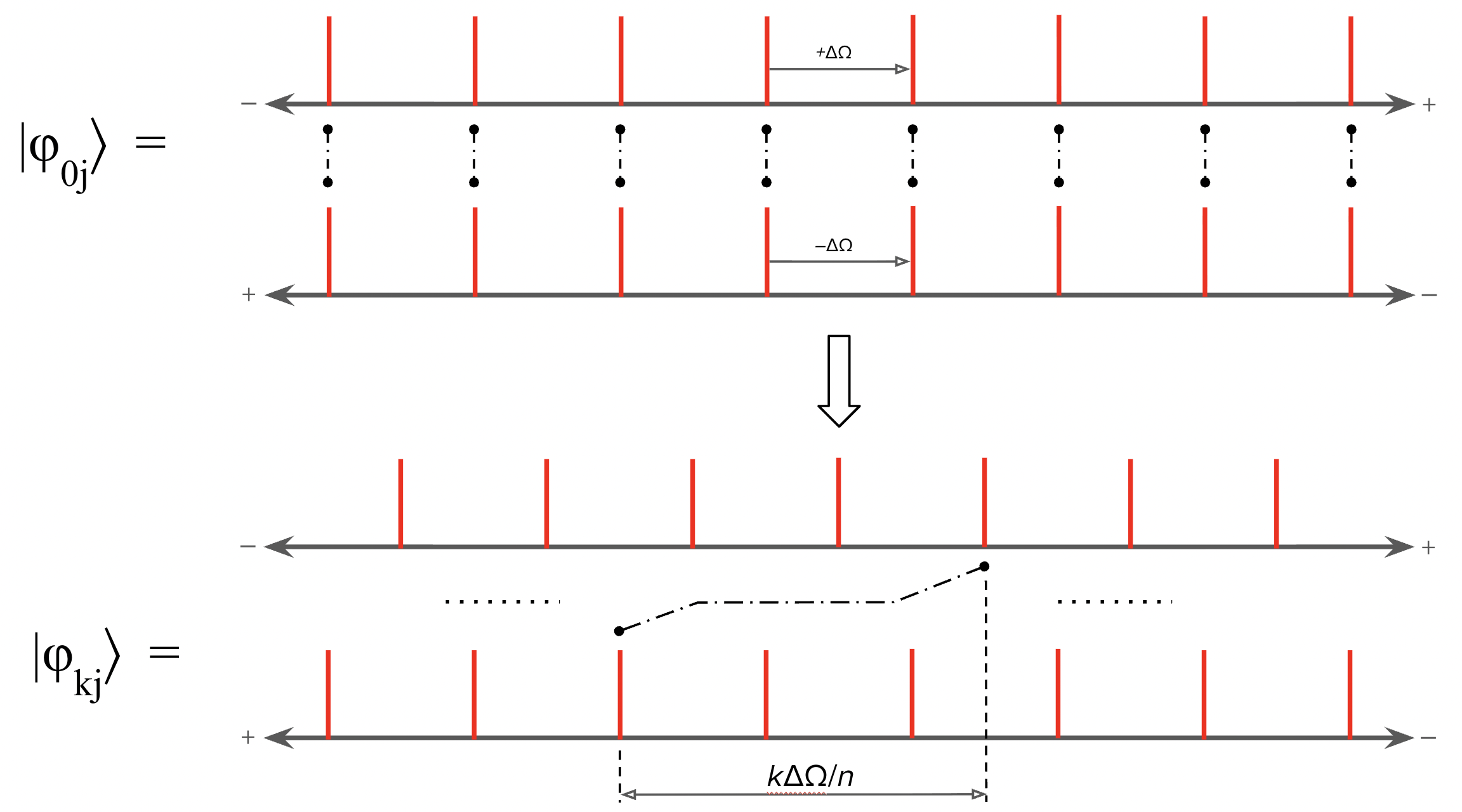}
    \caption{Schematic illustration of the second step of the encoding phase of our superdense coding protocol, described in \sect{sdc_encoding}, showing how the correlations in the biphoton frequency comb are affected by the frequency shift operator. In each panel, the top line represents photon $H$ and the bottom line represents photon $V$. Pairs of spikes linked by vertical dotted lines represent the pairs of frequency eigenstates of the form $\ket{\omega_1}^{\mathrm{freq}}\ket{\omega_2}^{
    \mathrm{freq}}$ that appear as summands in the EPR-like biphoton frequency comb. The top panel represents the state $\ket{\varphi_{0j}}$, before the frequency shift is applied ($\ket{\varphi_{0j}}$ includes a time-shift, but in the frequency basis this just applies a phase to each spike, so is not explicitly represented in the figure). The bottom panel represents the state $\ket{\varphi_{kj}}$ after a frequency shift, which acts on each summand as $\ket{\omega_1}^{\mathrm{freq}}\ket{\omega_2}^{
    \mathrm{freq}} \mapsto \ket{\omega_1 + \frac{k\DOmega}{n}}^{\mathrm{freq}}\ket{\omega_2}^{
    \mathrm{freq}}$, showing how arbitrary frequency shifts are distinguishable in the idealized case. In practice, the range of distinguishable shifts is limited by the available bandwidth.}
    \label{fig:freq_shift}
\end{figure}

Even though bandwidth, rather than the frequency basis comb spacing $\Delta\Omega$ from \eq{bfc_dirac}, is the relevant constraint, we will express the bin width as a fraction of $\Delta\Omega$ for analytical convenience. Let the frequency bin width be $\Delta\Omega/n$ for $n \in \mathbb{Z}$, implemented by the displacement operator $\fdisp{\Delta\Omega/n}$. Suppose Alice can apply up to $c$ distinguishable frequency shifts, where $c$ is determined by the bandwidth. Then for each $k \in \{0,\dots,c-1\}$, Alice can encode the $k{\text{th}}$ message by further applying the displacement operator
\begin{equation}
    \left(\fdisp{\frac{\Delta\Omega}{n}}\right)^k = \fdisp{\frac{k\Delta\Omega}{n}}
\end{equation}
to the state $\ket{\varphi_{0j}}$, to obtain
\begin{align}\label{eq:encoded_state}
 \ket{\varphi_{kj}} &\coloneq \fdisp{\frac{k\Delta\Omega}{n}}\ket{\varphi_{0j}}\\
 &\propto \sum_{m=-\infty}^{\infty} e^{-2\pi imj/d} \left|\frac{\omega_p}{2} + m\Delta\Omega + \frac{k}{n}\Delta\Omega\right\rangle_H^{\text{freq}} \left|\frac{\omega_p}{2} - m\Delta\Omega\right\rangle_V^{\text{freq}},
\end{align}
which is illustrated schematically in \fig{freq_shift}.

This concludes the encoding stage. Alice now transmits her photon to Bob, who upon possession of the two-photon state $\ket{\varphi_{kj}}$, must decode Alice's message by determining the values of $k$ and $j$.

As an aside, one might ask, at this point, why we are using a (biphoton) frequency comb to encode information in the first place, given that the frequency-domain encoding does not utilize the comb-like structure in an essential way, being bandwidth-limited rather than period-limited. The reason is that the comb structure allows us to encode further information in the conjugate variable, namely the temporal domain, via the temporal shift. 

\subsection{Decoding}\label{sec:sdc_decoding}
The decoding process operates on an encoded state in \eq{encoded_state} to recover the values of $k$ and $j$.

First, to measure the frequency shift and recover the value of $k$, we use a frequency beamsplitter (henceforth FBS). It is a two-photon operator defined to act on frequency eigenstates as follows:
\begin{equation}
    \ket{\omega_1}^{\text{freq}}\ket{\omega_2}^{\text{freq}} \xrightarrow{\text{FBS}}  \ket{\frac{\omega_1+\omega_2}{\sqrt{2}}}^{\text{freq}}\ket{\frac{\omega_1-\omega_2}{\sqrt{2}}}^{\text{freq}}
\end{equation}
with its operation on more general states following by linearity. The FBS can be thought of as a continuous-variable analog of a symmetric CNOT gate acting on the frequency variable.

The encoded state is first passed through an FBS, resulting in the following transformation (up to overall phases and constants):
\begin{align}\label{eq:after_FBS}
\ket{\varphi_{kj}} \xrightarrow{\text{FBS}} \sum_{m=-\infty}^{\infty} e^{-2\pi imj/d} \left|\frac{1}{\sqrt{2}}\left(\omega_p + \frac{k}{n}\Delta\Omega\right)\right\rangle_H^{\text{freq}} \left|\frac{1}{\sqrt{2}}\left(2m\Delta\Omega + \frac{k}{n}\Delta\Omega\right)\right\rangle_V^{\text{freq}}
\end{align}

The state in \eq{after_FBS} is separable, so the two photons have been disentangled by the FBS. The horizontally polarized photon $H$ is now measured in the frequency basis. Its frequency, $\frac{1}{\sqrt{2}}\left(\frac{\omega_p}{2} + \frac{k}{n}\Delta\Omega\right)$, reveals the value of $k$. Here we see explicitly that in principle, any value of $k$ is recoverable by this procedure. Next, to recover the value of $j$, the vertically polarized photon $V$ is measured in the temporal basis. The state of $V$, initially expressed in the frequency basis, can be Fourier transformed and re-expressed in the temporal basis as:
\begin{align}
\ket{\varphi_V} &\propto \sum_{m=-\infty}^{\infty} e^{-2\pi imj/d} \left|\frac{1}{\sqrt{2}}\left(2m\Delta\Omega + \frac{k}{n}\Delta\Omega\right)\right\rangle_V^{\text{freq}} \\
&\propto \sum_{m=-\infty}^{\infty} e^{i\frac{T_m k \Delta\Omega}{n\sqrt{2}}} |T_m\rangle_V^{\text{time}}
\end{align}
where $T_m = \sqrt{2}\left(\frac{j}{d}\frac{\pi}{\Delta\Omega} + m\frac{\pi}{\Delta\Omega}\right)$. Since the value of $j$ is encoded in $T_m$, we measure the photon $V$ in the temporal basis. The quantum state, a superposition over $m$, collapses to yield $T_m$ for a uniformly sampled $m \in \mathbb{Z}$, allowing $j$ to be recovered from the value of $T_m$ if $0 \leq j < d$. Recovering $k$ and $j$ concludes decoding.

\subsection{Connection to physical states}\label{sec:bfc_envelopes}
The idealized biphoton comb states used in \sect{sdc_protocol}, being infinitely squeezed, are unphysical. To relate this calculation to a more physical model, we modify the initial state $\ket{\varphi_{00}}$ in two ways \citep{Yamazaki2023}. First, to represent the finite linewidth of the frequency spikes, the Dirac comb is convolved with a function $f(\Omega)$ representing the lineshape. Second, we take the pointwise product of the convolution with an envelope function $g(\Omega)$, to represent the finite bandwidth. The resulting quantum state reproduces \eq{bfc_exp_freq} (representing the experimentally prepared state, described in \sect{exp_generation}):
\begin{align}\label{eq:comb_convolved}
\ket{\varphi_{00}} = \int_{-\infty}^{\infty} d\Omega\ g(\Omega)\left[\left(\sum_{m=-\infty}^{\infty} \delta(\Omega - m\Delta\Omega)\right) * f(\Omega)\right] \hat{a}_{H}^{\dagger}\left(\frac{\omega_p}{2} + \Omega\right) \hat{a}_{V}^{\dagger}\left(\frac{\omega_p}{2} - \Omega\right) \ket{\text{v.s.}}
\end{align}

Since the Fourier transform exchanges convolutions for pointwise products and vice versa, the time-domain state has an analogous structure. A frequency shift causes both $f(\Omega)$ and $g(\Omega)$ to shift, and temporal shifts behave analogously. A frequency shift induces translations of the pulse envelope $g(\Omega)$ and the ideal frequency comb. Likewise, temporal shifts induce translations of the temporal envelope, and the ideal temporal comb. This structure ensures proper encoding and decoding on the physical state in \eq{comb_convolved}. In next section, we elaborate on the physical implementation of our protocol by explicitly analyzing an experimentally viable biphoton comb state and noise model.

\section{Experimental design}\label{sec:exp_design}
\subsection{Generation of biphoton frequency combs}\label{sec:exp_generation}
We can create biphoton frequency combs in a type-II spontaneous parametric down-conversion (SPDC) configuration. $\Delta\Omega$ is the free-spectral range (FSR) of the cavity. $\Omega$ refers to the detuning of biphotons from frequency degeneracy at half the pump frequency $\omega_p$. Unlike the Dirac comb, as mentioned in \sect{bfc_envelopes}, in practice comb-like patterns with Lorentzian cavity lineshapes appear in the frequency-domains \citep{Xie2015,Chang2021,Chang2025a, Chang2023}. Hence, the frequency basis of the entangled TFGKP state is determined by the quantum frequency combs created through photons with central frequencies that have been shifted.
\begin{align}\label{eq:bfc_exp_freq}
\ket{\varphi} = \sum_{m=-N_0}^{N_0} \int d\Omega\ g(\Omega)f(\Omega - m\Delta\Omega)\hat{a}_{H}^{\dagger}\left(\frac{\omega_p}{2} + \Omega\right) \hat{a}_{V}^{\dagger}\left(\frac{\omega_p}{2} - \Omega\right) \ket{\text{v.s.}}
\end{align}

$2N_0 + 1$ is the number of cavity lines passed by a bandwidth-limiting filter; $g(\Omega) = \text{sinc}(A\Omega)$ is the phase-matching function of the biphoton source, where $A = 2.78/\pi B_{PM}$ with $B_{PM}$ being the full-width-half maximum (FWHM) bandwidth; $f(\Omega - m\Delta\Omega)$ is the cavity's Lorentzian transmission lineshape with FWHM linewidth of $2\Delta\omega$; and $\Delta\Omega$ is the cavity FSR. The temporal-basis representation of the biphoton frequency comb can be established through the Fourier transformation of \eq{bfc_exp_freq}:
\begin{align}\label{eq:bfc_exp_time}
\ket{\varphi} = \int dt \int d\tau\  e^{-i\frac{\omega_p}{2}(2t-\tau)} e^{-\Delta\omega|\tau|} \sum_{m=-N_0}^{N_0} \text{sinc}(A m \Delta\Omega) \cos(m\Delta\Omega\tau)\hat{a}_{H}^{\dagger}(t)\hat{a}_{V}^{\dagger}(t - \tau)\ket{\text{v.s.}},
\end{align} 
where we have used $\Delta\Omega/2\pi \ll B_{PM}$. Hence, the temporal-basis states exhibit numerous discretized peaks in the relative time delay $\tau$, with a repetition period equal to the cavity round-trip time, $\Delta T = 2\pi/\Delta\Omega$. The frequency-bin information remains resilient to incoherent spectral broadening since each frequency bin is spectrally isolated under the assumption of $\Delta\Omega \gg \Delta\omega$, which is typically satisfied in experiments \citep{Xie2015,Ikuta2019,Chang2021,Chang2025a,Chang2023}. Conversely, in the case of temporal-basis information, the finite temporal resolution of the detector introduces considerable incoherent temporal broadening. Such temporal broadening can be circumvented with lower-jitter single-photon detectors \citep{Chang2023b}, or ultranarrow-band cavity FSR \citep{Chang2024b}.

\begin{figure}
    \centering
    \includegraphics[width=\columnwidth]{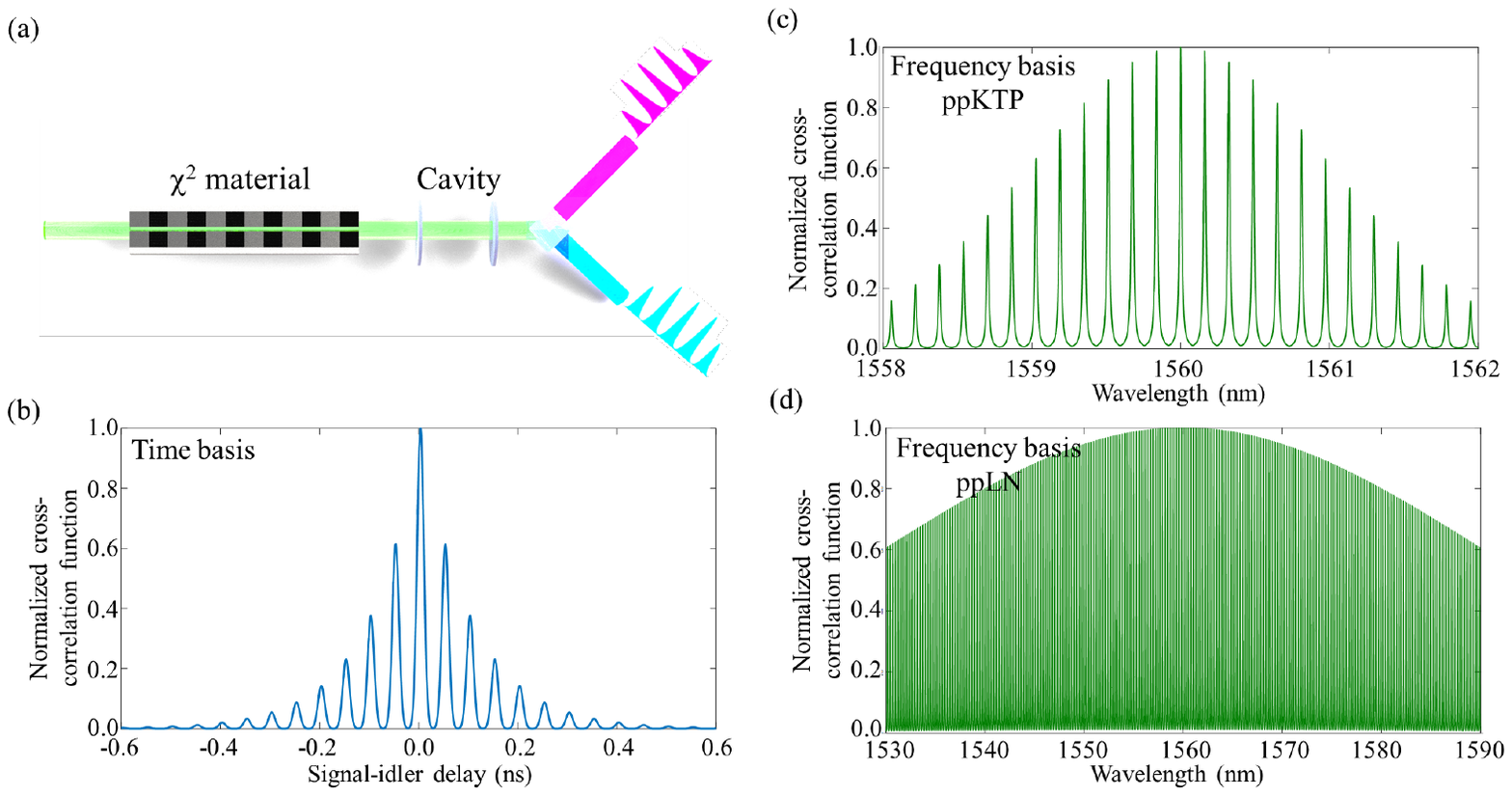}
    \caption{Modeling the probability distributions of a biphoton frequency comb (EPR-like entangled TFGKP state) in both the time and frequency domains. (a) Schematic of a biphoton frequency comb generation. (b) The signal-idler temporal correlation in the comb state with a 20 GHz FSR cavity. Frequency spectrum of the comb state, with phase-matching FWHM bandwidth of (c) 250 GHz and (d) 7.4 THz.}
    \label{fig:bfc_states}
\end{figure}

Fig.~\ref{fig:bfc_states} displays instances of tangible biphoton frequency comb (BFC) states. Fig.~\ref{fig:bfc_states}(a) is a BFC state via cavity-assisted scheme \citep{Xie2015,Chang2021}. We illustrate the time and frequency characteristics of the BFC state via a periodically-poled KTiOPO4 (ppKTP) waveguide in Figs.~\ref{fig:bfc_states}(b) and~\ref{fig:bfc_states}(c). Fig.~\ref{fig:bfc_states}(b) is the time basis for a BFC state with a 20 GHz FSR cavity, a 2 GHz linewidth, and an FWHM timing-jitter of 20 ps. The jitter is chosen to resolve the temporal discretized peaks of a BFC state, and these temporal correlations of multiple peaks supports the phase-coherence among different frequency bins \citep{Ikuta2019,Chang2023,Chang2024b}. In Fig.~\ref{fig:bfc_states}(c), we plot the frequency basis of the same BFC state, with phase-matching FWHM bandwidth of 250 GHz. We see that in a frequency basis, our BFC state comprises both Gaussian and Lorentzian lineshapes. From Eqs.~\ref{eq:bfc_exp_freq} and~\ref{eq:bfc_exp_time}, we note that based on the Fourier duality, the superposition of all the discretized spectral peaks in the quantum frequency comb results in the discretization of temporal peaks, displaying oscillations in the correlation function at a period equal to the inverse of the frequency bin spacing, $\Delta T = 2\pi/\Delta\Omega$ \citep{Chang2023}. These results match Fig.~\ref{fig:bfc_states}(b) and~\ref{fig:bfc_states}(c). Then, in Fig.~\ref{fig:bfc_states}(d), we illustrate the BFC state using a broadband periodically poled lithium niobate (ppLN) waveguide with a FWHM bandwidth of 7.4 THz. For all calculations in Fig.~\ref{fig:bfc_states}, the center wavelength of the biphoton source is set at 1560 nm. Additionally, \app{app_b} contains a different BFC state (singly-filtered configuration) \footnote{\app{app_b} includes the biphoton frequency comb based on singly-filtered configuration, proposed experimental configuration of singly-filtered biphoton frequency comb for superdense coding protocol with FBS operation, and experimental configurations of biphoton frequency comb for superdense coding with frequency-polarization hyperentanglement.}.

\subsection{Near-deterministic frequency beamsplitter operation}\label{sec:exp_fbs}
\begin{figure}
    \centering
    \includegraphics[width=0.5\columnwidth]{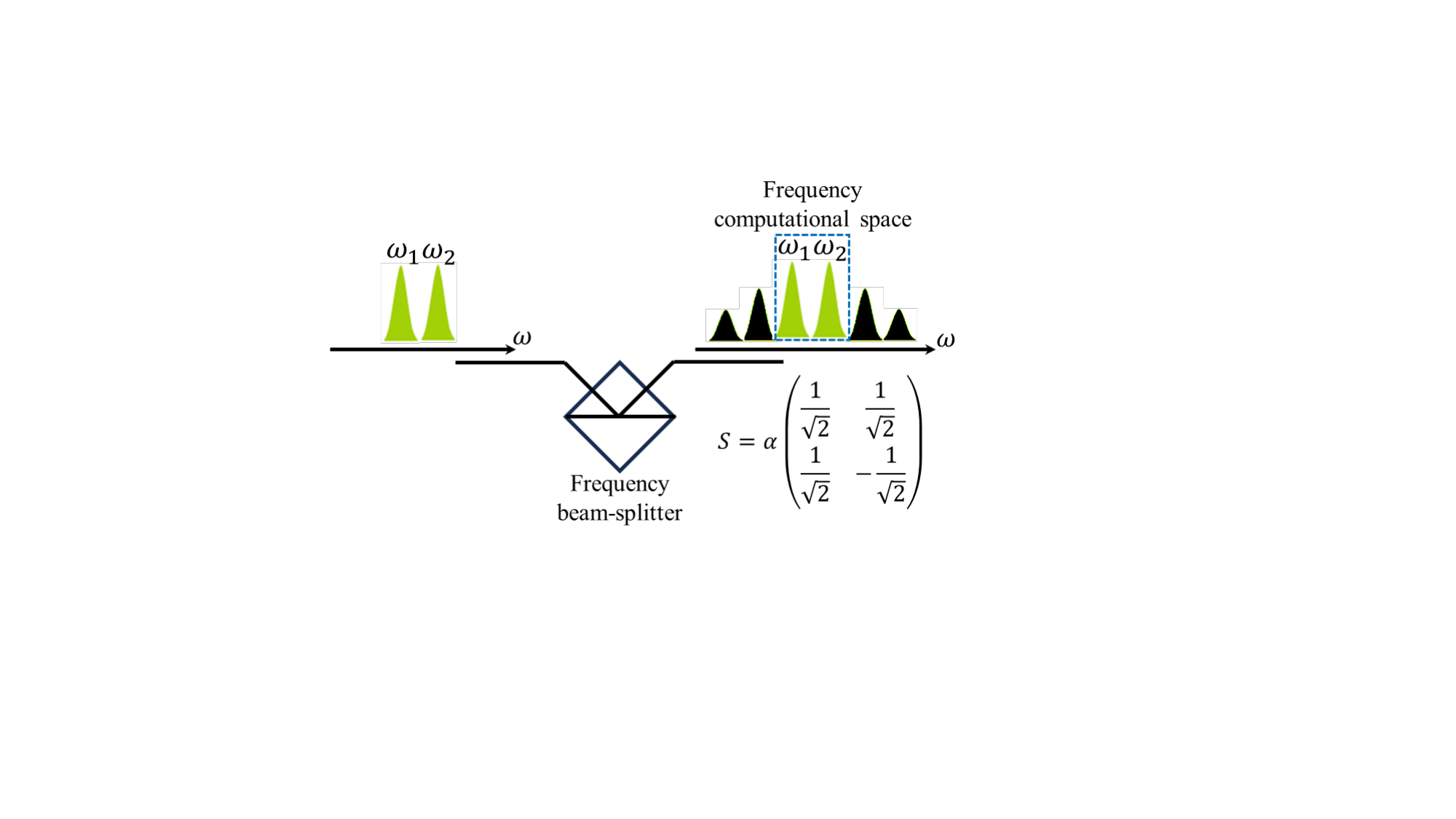}
    \caption{Frequency beamsplitter (FBS) operation via a phase modulator. $\omega_1$ and $\omega_2$ are the two input and output spectral modes that belong to the computational-space of FBS operation. The dark green frequency bins at the output of FBS indicate the undesired sideband population, causing this scheme to be probabilistic \citep{Imany2018}. Here we consider the sandwiched scheme \citep{Lu2018,Seshadri2022} for near-deterministic FBS operation. The optical loss of our proposed scheme can be 5 dB \citep{Nussbaum2022}.}
    \label{fig:fbs_operation}
\end{figure}

We first introduce the near-deterministic FBS operation, which is crucial to the decoding process for superdense coding. Previously proposed versions of FBS were demonstrated based on nonlinear frequency conversion \citep{Kobayashi2016,Fabre2022a,Fabre2022b} and linear optics \citep{Imany2018}. Here we choose a linear optics-based FBS because the scheme does not contain a noise source and its performance does not suffer from multiphoton components. In Fig.~\ref{fig:fbs_operation}, we illustrate the implementation of FBS via a phase modulator. $\omega_1$ and $\omega_2$ are two input spectral modes within the computational-space. We represent the scattering matrix of this FBS operation, which is similar to conventional spatial BS, where $\alpha$ represents settings for transitioning to alternative frequency modes beyond the computational-space. If both input photons are either transmitted or reflected by FBS, the relative phase between these two events is $\pi$. The dark green frequency bins at the output of FBS indicate the undesired sideband population, causing this scheme to be probabilistic \citep{Imany2018}. The probabilistic nature of FBS can be overcome by involving two EOMs with a pulse shaper positioned between them \citep{Lu2018,Seshadri2022}. The spectral phase introduced by the intermediate stage guarantees that the sidebands generated following the initial EOM are reintegrated into the computational-space. This arrangement enables a near-deterministic FBS \citep{Lu2018,Seshadri2022}. The whole sandwiched structure typically has a total insertion loss of about 12 dB \citep{Lu2018,Lu2023}. However, recent advances in thin-film lithium niobate provide promising opportunities for achieving high-speed integrated EOMs with lower optical loss (< 1 dB) \citep{Wang2018,Ren2019}. Besides lower insertion loss EOMs, a recent study suggests that together with on-chip pulse shapers, this integrated sandwiched structure can have only around 5 dB total loss \citep{Nussbaum2022}. Considering the typical loss in Bell-state measurement of 6 dB from linear optics \citep{Barreiro2008,Williams2017,Hu2018,Calsamiglia2001,Kwiat1998,Wei2007,Wang2019}, this makes the FBS scheme proposed in this work more attractive in high-rate superdense coding protocol.

\subsection{Encoding and decoding}\label{sec:exp_encdec}
\begin{figure}
    \centering
    \includegraphics[width=\columnwidth]{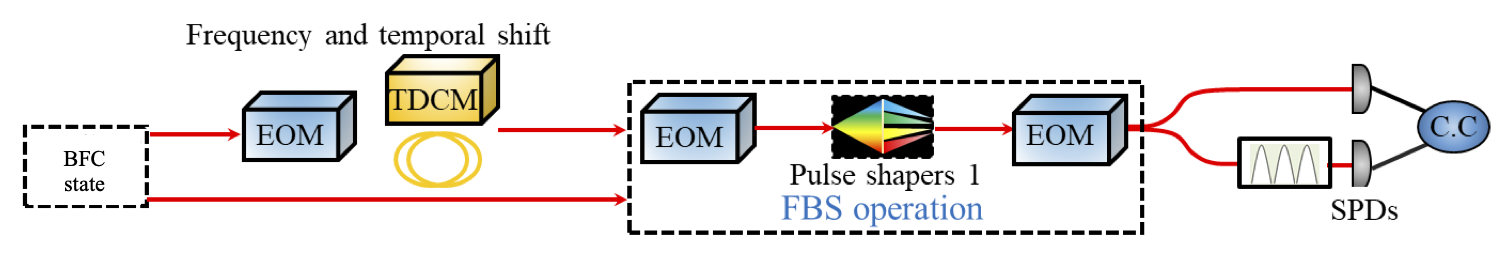}
    \caption{Proposed experimental setup for implementation of superdense coding with biphoton frequency combs and FBS operation. The temporal shifts can be provided via TDCM or telecom-fiber with different delays. C.C., coincidence counts. An additional tunable-filter is for frequency measurements to conclude the superdense coding protocols via two SPDs.}
    \label{fig:experimental_setup}
\end{figure}

Now we explain how our superdense coding protocol can be implemented in practice, as in Fig.~\ref{fig:experimental_setup}. We first generate biphoton frequency comb states using a fiber cavity and a SPDC source. The first EOM provides frequency shifts for pump wavelength, which can then be decoded using an FBS operation \citep{Seshadri2022} and frequency measurement. The temporal shift can be realized with either a tunable dispersion compensation module (TDCM) \citep{Franson1992,Mower2013,Chang2024a} or telecom-fibers of different lengths after the first EOM. A near-deterministic FBS operation can be realized using the sandwiched scheme \citep{Lu2018,Seshadri2022,Lu2023}. All the EOMs are synchronized with a single radio-frequency source. Following the FBS operation, we incorporate an extra tunable frequency-filter to conduct frequency-basis measurements for one of the photons, whereas the second photon is directed towards a single-photon detector (SPD) for temporal-basis measurements. This determines the temporal and spectral shifts applied to the initial state, distinguishing between different possible encoded states. \sect{sfc_sdc} of \app{app_b} contains the experimental design for superdense coding with a singly-filtered biphoton frequency comb. Additionally, \sect{fp_sdc} of \app{app_b} contains a superdense coding design using a frequency-polarization hyperentanglement scheme. We comment that this alternative scheme has similar channel capacity compared to the FBS-based superdense coding protocol proposed here. The advantage of the hyperentanglement scheme is it requires mostly passive telecommunication components, such as standard frequency and polarization filters.

\section{Channel capacity of proposed superdense coding protocol}\label{sec:channel_capacity}
Our next task is to provide an estimate of the transmission capacity of our protocol given realistic experimental parameters. As discussed in \sect{sdc_protocol}, the upper limit on the transmission capacity in superdense coding is dictated by the number of distinguishable states \citep{Harrow2004,Barreiro2008,Williams2017,Hu2018}. This amounts to upper bounding the number of distinguishable frequency and temporal shifts, denoted $c$ and $d$ respectively. In the idealized protocol described in \sect{sdc_protocol}, shifts by distinct amounts are perfectly distinguishable. However, the finite precision of experimental operations introduces errors in the distinguishability of distinct shifts, which would compromise the accuracy of encoding and decoding. This section explains how to incorporate experimental errors to compute a realistic capacity estimate for our protocol, and is structured as follows.

In \sect{exp_params}, we specify the sources and numerical values of these errors for current state-of-the-art hardware, and state the resulting transmission rates of our scheme, whose computation we explain in the next section. In \sect{capacity_calc}, we explain how applying classical error correction to the frequency and temporal shifts (by viewing the procedure as a noisy channel) allows for an asymptotically vanishing error rate. We find that the resulting error-corrected channel capacity of our superdense coding protocol is 3.73 bits per transmitted photon for the ppKTP parameters and 8.91 bits per transmitted photon for the ppLN parameters.

Finally, we make three natural comparisons of our protocol's transmission capacity to that of other ways of transmitting classical information using quantum states. We compare to single-photon frequency combs in \sect{single_comb_comp}; to the qudit version of textbook superdense coding implemented with logical Bell states of TFGKP qudits in \sect{gkp_bell_comp}; and to prior schemes from the superdense coding literature in \sect{prior_sdc_comp}. Our protocol improves upon all of these methods, and surpasses the previously highest achieved superdense coding capacity (4 bits in the Kwiat-Weinfurter scheme) by a factor of 2.2. All of these comparisons assume the ppLN parameters for our protocol, with capacity 8.91 bits.

\subsection{Relevant experimental parameters}\label{sec:exp_params}

\subsubsection{Frequency operations}
As discussed in \sect{sdc_protocol}, arbitrary frequency shifts applied to our initial biphoton comb state are distinguishable in principle. In practice, the number of frequency shifts is limited by available bandwidth and frequency bin size, which is determined by the finite resolution of shifts and measurements. Considering the precision of the frequency shift $\sigma_{f,shift}$ to be 10 GHz, and the resolution of frequency measurement $\sigma_{f,meas}$ to be 10 GHz, we treat them as Gaussian standard deviations. The total standard deviation from both sources of error is then $\sigma_{f,total} = \sqrt{\sigma_{f,shift}^2 + \sigma_{f,meas}^2} = 10\sqrt{2}$ GHz. Note that these parameters also implicitly include the effect of finite linewidth in a Gaussian approximation. To achieve a raw error rate of 1\%, we choose the frequency bin size to be $2.58 \times \sigma_{f,total}$, where the constant is calculated using the complementary error function. The number of distinguishable frequency states is $c = \frac{FWHM\text{ bandwidth}}{2.58 \times \sigma_{f,total}}$. Given the parameters of comb states in \fig{bfc_states}, for ppKTP FWHM bandwidth of 250 GHz, $c$ is 6 states; and for ppLN FWHM bandwidth of 7.4 THz, $c$ is 202 states. To minimize the probability of error to zero, one must use fewer states or implement an error-correcting code. The latter is more efficient, as we explain below. Optimizing the ppLN frequency-basis transmission capacity over the number of frequency bins using Shannon's noisy coding theorem \citep{Cover2006} yields 6.98 bits at vanishing error, which corresponds to 126 distinguishable states per transmitted photon, in the frequency channel. We explain how to compute this capacity in \sect{capacity_calc}.

\subsubsection{Temporal operations}
The temporal-basis measurement, in the second step of the decoding procedure, behaves differently from the frequency measurement, since the biphoton comb is an eigenstate of temporal shifts by an integer multiple of the temporal-domain comb period $\Delta T = 2\pi/\Delta\Omega$. The upper limit $d$ is the number of distinguishable states between adjacent spikes of the comb. The key parameters are the temporal-domain comb period $\Delta T$, the precision of the temporal shift $\sigma_{t,shift}$, the precision of the photon arrival-time measurement $\sigma_{t,meas}$, and the time-domain linewidth of SPDC source $\sigma_{linewidth}$. We estimate the parameter $\sigma_{linewidth}$, depending on the FWHM bandwidth, so we see that the biphoton source bandwidth affects both frequency and time-domain channel capacities by controlling the time-domain spike linewidth. Analogously to the frequency domain, with the same Gaussian approximation, the total standard deviation $\sigma_{t,total} = \sqrt{\sigma_{t,shift}^2 + \sigma_{t,meas}^2 + \sigma_{linewidth}^2}$. Given $\sigma_{t,shift}$ of 1 ps, $\sigma_{t,meas}$ of 3 ps, ppKTP FWHM linewidth $\sigma_{linewidth}$ of $\approx 7.1$ ps, $\sigma_{t,total}$ is 7.8 ps. The temporal period of the comb state $\Delta T$ is $\approx 50$ ps. Binning the arrival times leads to a noisy channel for the temporal-shift information with capacity 1.63 bits per transmitted photon, for the ppKTP parameters. For ppLN FWHM bandwidth of 7.4 THz, $\sigma_{linewidth} \approx 0.24$ ps, the $\sigma_{t,total}$ is 3.17 ps. This leads to a higher capacity: 1.93 bits per transmitted photon, in the temporal channel.

The total capacity is the sum of the capacities of the frequency and temporal channels.

\subsubsection{Achieving experimental parameters with current technologies}
Before moving on to the details of the capacity calculations, we summarize how the experimental parameters for maximum transmission capacity in our protocol can be achieved with current technologies, focusing on the main errors occurring during the biphoton comb generation and measurements. The consistent spectral broadening is characterized by a Lorentzian distribution, mainly affected by the spectrum of the cavity in state preparation (although we use a Gaussian approximation in our capacity calculations for analytical simplicity). We chose for the purposes of illustration to compare the transmission capacities of ppKTP and ppLN sources, with 250 GHz and 7.4 THz FWHM bandwidths respectively, which are commercially available \citep{Xie2015,Chang2021,Chang2025a,Chang2023}. For precision of frequency shift $\sigma_{f,shift}$, and resolution of frequency measurement $\sigma_{f,meas}$ of 10 GHz, these numbers are available from standard EOM and tunable-filters \citep{Ikuta2019,Chang2021,Chang2025a,Chang2023,Seshadri2022}. For the temporal domain, we choose to use a 20 GHz FSR to ensure a reasonable number of distinguishable states under reasonable errors. Meanwhile, the temporal shift $\sigma_{t,shift}$ of 1 ps is possible given the temporal resolution of tunable fiber delay-lines and commercial tunable dispersion modules. Additionally, the precision of the photon arrival-time measurement $\sigma_{t,meas}$ of 3 ps is based on recent low-jitter single-photon detectors \citep{Korzh2020}. Here the corresponding temporal spacing $\Delta T = 2\pi/\Delta\Omega$ is larger than $T_{jitter}$, where $T_{jitter}$ is the effective temporal resolution of low-jitter detectors \citep{Jaramillo-Villegas2017,Ikuta2019,Imany2019,Reimer2019,Maltese2020,Chang2021,Chang2025a}. Therefore, according to our proposed approaches, the most advanced technologies presently fulfill the experimental prerequisites for superdense coding utilizing entangled TFGKP states.

\subsection{Calculation of error-corrected channel capacity}\label{sec:capacity_calc}
Now we provide additional details on computing the error-corrected transmission rates mentioned above. As explained previously, in our protocol, the discretization of the continuous frequency and time parameters leads to a finite set of possible messages that can be communicated with each photon transmission, corresponding to the number of distinguishable photonic states: $c$ states in the frequency domain and $d$ states in the temporal domain. Considering both domains, the total number of distinguishable states is $cd$, corresponding to a transmission rate of $\log_2 c + \log_2 d$ bits per transmitted photon. In practice, these states are not perfectly distinguishable, due to the error sources discussed above. The upshot is that, due to the finite precision of hardware operations such as shifts and measurements (in both frequency and time domains), as well as the effect of finite linewidth (in the time domain), the received message may not always be the one intended to be transmitted by the sender.

There are now two ways to proceed. The first is to choose the size of the discretized frequency and time bins such that the error probability is a sufficiently small constant. An example provided earlier in this section is that an error rate of 1\% is achievable with 202 distinguishable states in the frequency domain. This is a hardware-based approach to reducing the error rate. The second option is to minimize the error rate to zero (asymptotically) by adopting a software-based solution: viewing the photon transmission as communication over a noisy channel and using an error-correcting code. In this case, each frequency or time bin alone does not correspond to a distinct message. Rather, the sender encodes the message according to an error-correcting code, so that at the receiver's end, the map from received frequency or time bins to distinct messages is many-to-one. This principle, of software-level redundancy providing resilience against hardware-level errors, is the information-theoretic basis for reliable communication over noisy channels \citep{Cover2006}. Shannon's noisy channel coding theorem enables us to calculate the highest achievable number of bits per transmitted photon as the channel capacity of our photonic channel \citep{Cover2006}. Below, we explain how these channel capacities were computed.

Per the discussion of errors above, we use a Gaussian noise model for both the frequency and time channels, and calculate the capacity of each. Starting with frequency, suppose the intended frequency shift is $f$, and let the frequency shift measured by the receiver be $g$. Due to the finite precision of frequency shifts and measurements, the probability density for $g$ is a Gaussian centered at $f$:
\begin{align}\label{eq:gauss_noise_model}
P(g) = \frac{1}{\sqrt{2\pi\sigma^2}}e^{-(g-f)^2/(2\sigma^2)}
\end{align}

We take the standard deviation $\sigma = 10\sqrt{2}$ GHz based on experimentally feasible parameters. Recall that the range of allowed frequency shifts (limited by the bandwidth) is discretized into $N$ bins. These $N$ bins are the symbols that constitute the alphabet for our noisy channel. To specify the channel, we must provide the conditional probability that the receiver measures bin $y$, given that sender intended to transmit bin $x$, where any bin $i$ is indexed by its midpoint frequency, $f_i$. We can calculate these conditional probabilities using \eq{gauss_noise_model}. Let the size of each bin be $\Delta$ GHz. Then the relevant conditional probabilities, collectively constituting the channel's transition matrix, are given by:
\begin{align}\label{eq:transition_probs}
P(\text{output bin} = y|\text{input bin} = x) = \int_{f_y - \Delta/2}^{f_y + \Delta/2} ds\ \frac{1}{\sqrt{2\pi\sigma^2}}e^{-(s-f_x)^2/(2\sigma^2)}
\end{align}

Evidently each bin is most likely to be mapped under noise to nearby bins on either side, with the probability of mapping to faraway bins decreasing rapidly according to the Gaussian. Since the bandwidth we use is finite, the frequencies near the edge of the bandwidth can be mapped to frequencies outside the bandwidth. For analytical simplicity, we approximate our channel as having periodic boundary conditions, which allows us to neglect this edge effect and results in a symmetric channel. By Shannon's noisy channel coding theorem, the channel capacity $C$ is given by the maximum mutual information between the channel's input and output, maximized over all probability distributions over the input symbols. Denoting the input and output by random variables $X$ and $Y$ respectively:
\begin{align}
C = \max_{p(X)}I(X:Y) = \max_{p(X)}[H(X) - H(X|Y)] = \max_{p(X)}[H(Y) - H(Y|X)]
\end{align}

Due to the symmetry of the channel, the maximizing input distribution is the uniform distribution, with $p(X = x) = \frac{1}{N}$ for all $x$. Again by symmetry, the corresponding distribution of the output $Y$ is uniform: $p(Y = y) = \frac{1}{N}$ for all $y$. So we can express the capacity as $C = \log N - H(Y|X)$. The conditional entropy $H(Y|X)$ can be computed using the transition matrix defined by \eq{transition_probs} to complete the capacity calculation.

\begin{figure}
    \centering
    \includegraphics[width=0.9\columnwidth]{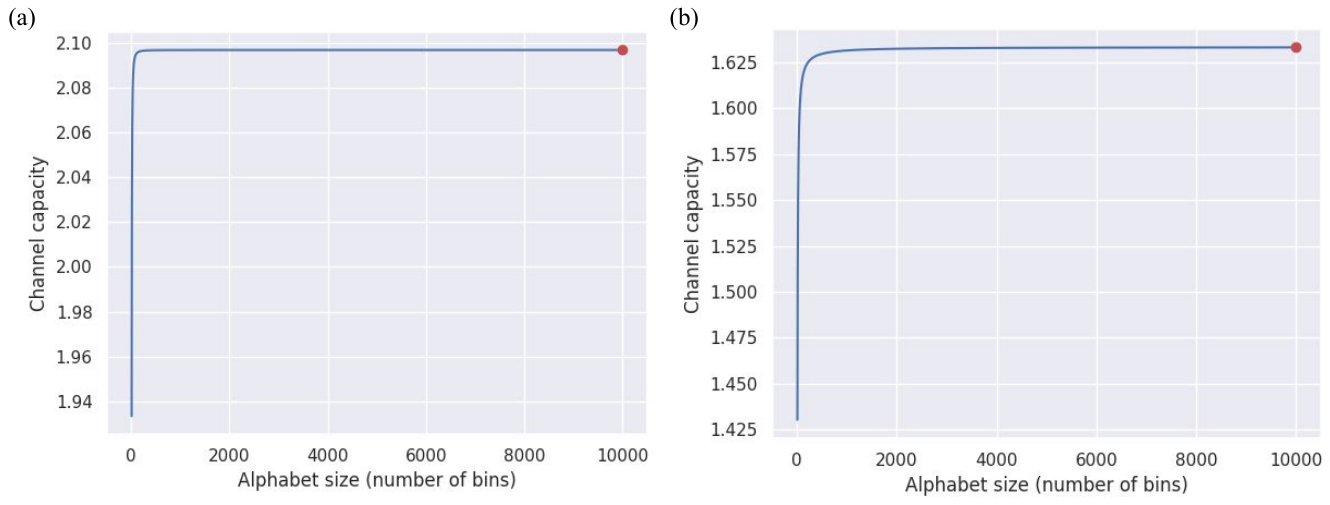}
    \caption{Calculated superdense coding channel capacity of biphoton frequency comb state in ppKTP. (a) Frequency-basis channel capacity as a function of number of bins, for ppKTP FWHM bandwidth = 250 GHz, frequency shift precision $\sigma_{f,shift}$ = 10 GHz, and frequency measurement resolution $\sigma_{f,meas}$ = 10 GHz. The asymptotic channel capacity in the frequency basis is 2.1 bits/photon. (b) Temporal-basis channel capacity as a function of number of bins, for ppKTP FWHM bandwidth = 250 GHz, time-domain comb period $\Delta T$ = 50 ps, temporal shift precision $\sigma_{t,shift}$ = 1 ps, photon arrival-time measurement precision $\sigma_{t,meas}$ = 3 ps, and time-domain linewidth of SPDC source $\sigma_{linewidth}$ = 7.1 ps. The asymptotic channel capacity in the temporal basis is 1.63 bits/photon.}
    \label{fig:capacity_ppktp}
\end{figure}

\begin{figure}
    \centering
    \includegraphics[width=0.9\columnwidth]{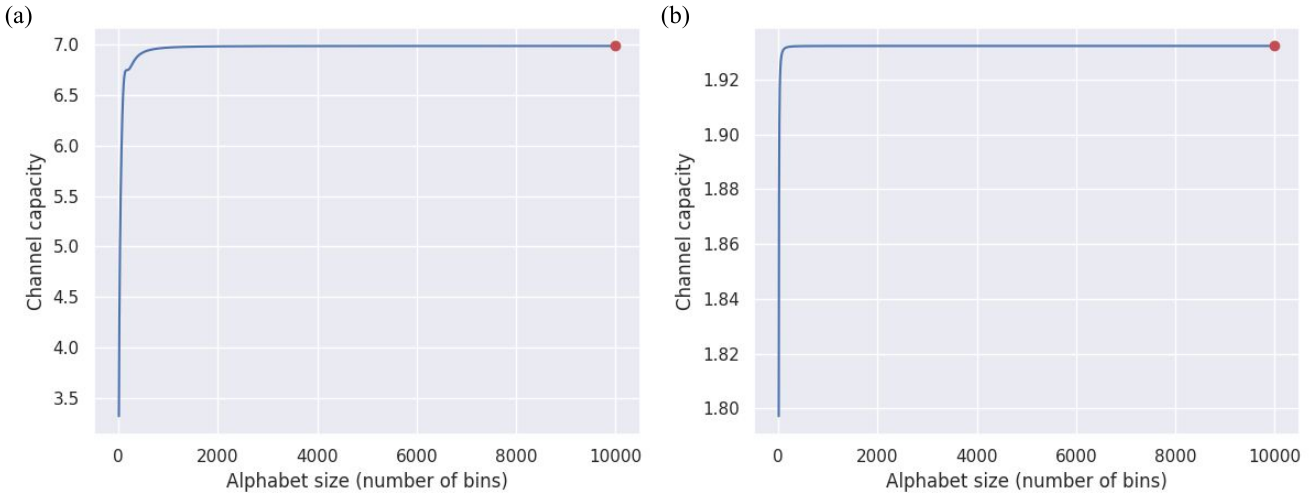}
    \caption{Calculated superdense coding channel capacity of biphoton frequency comb state in ppLN. (a) Frequency-basis channel capacity as a function of number of bins, for ppLN FWHM bandwidth = 7.4 THz, frequency shift precision $\sigma_{f,shift}$ = 10 GHz, and frequency measurement resolution $\sigma_{f,meas}$ = 10 GHz. The asymptotic channel capacity in the frequency basis is 6.98 bits/photon. (b) Temporal-basis channel capacity as a function of number of bins, for ppLN FWHM bandwidth = 7.4 THz, time-domain comb period $\Delta T$ = 50 ps, temporal shift precision $\sigma_{t,shift}$ = 1 ps, photon arrival-time measurement precision $\sigma_{t,meas}$ = 3 ps, and time-domain linewidth of SPDC source $\sigma_{linewidth}$ = 0.24 ps. The asymptotic channel capacity in the temporal basis is 1.93 bits/photon. Hence, the maximum transmission rate for our superdense coding scheme is the sum of the capacities of the frequency and temporal channels with the ppLN, which is up to 8.91 bits/transmitted photon.}
    \label{fig:capacity_ppln}
\end{figure}

We numerically calculate the frequency-basis capacity for a range of values of the alphabet size $N$, starting with $N = 10$, for both the ppKTP and ppLN parameters (see Fig.~\ref{fig:capacity_ppktp} and Fig.~\ref{fig:capacity_ppln} for the calculation results). After a rapid initial increase in both cases, the frequency-basis capacity appears to asymptote to 2.1 bits for ppKTP and 6.98 bits for ppLN. For the temporal basis, we use an analogous discrete channel with Gaussian noise. In this case, periodic boundary conditions are exact since the photon being measured is in a comb state, as shown in our calculations above. The capacity calculation is analogous to the frequency case. Here too, the capacity increases with the alphabet size, and appears to asymptote to 1.63 bits for ppKTP and 1.93 bits for ppLN. In Fig.~\ref{fig:capacity_ppktp}, we plot these capacities for alphabet sizes ranging from $N = 10$ to $N = 10^4$, using ppKTP parameters. Then, in Fig.~\ref{fig:capacity_ppln}, we plot these capacities for alphabet sizes ranging from $N = 10$ to $N = 10^4$ with ppLN parameters. The resulting maximum transmission rate for our superdense coding scheme is the sum of the capacities of the frequency and temporal channels with the ppLN parameters: 6.98 bits per transmitted photon in the frequency basis and 1.93 bits per transmitted photon in the temporal basis. The estimated maximum total channel capacity of our superdense coding scheme is therefore 8.91 bits per transmitted photon, corresponding to 481 distinguishable messages.

\subsection{Comparison to other communication protocols}\label{sec:comparisons}

\subsubsection{Comparison to single-photon frequency comb}\label{sec:single_comb_comp}
In the previous section, we separately computed the number of error-corrected bits transmitted through the frequency channel and the temporal channel in our protocol. Since our biphoton frequency comb is an eigenstate of temporal shifts representing integer multiples of $2\pi/\DOmega$, the aforementioned temporal channel capacity is equivalent to the maximum number of error-corrected bits that can be encoded by temporal modulation of a single-photon comb of period $2\pi/\DOmega$. We found the answer to be 1.93 bits. 

Therefore, our estimated channel capacity of 8.91 bits per transmitted photon is 4.6 times the 1.93 bits per transmitted photon achievable using temporal modulation of a single-photon frequency comb with the same parameters. 

By an analogous calculation, we could also compare our rate to that achievable with frequency modulation of a single-photon comb with the same parameters, where the number of distinguishable states is now limited by the frequency-domain comb period. However, the capacity of a single-photon comb with frequency modulation turns out to be several orders of magnitude lower than what is achieved by our biphoton frequency comb with the same parameters. (Recall that in the biphoton case, the frequency channel capacity is bandwidth-limited rather than period-limited, due to the state's entanglement structure). So to provide a fair comparison, we compare to the rate achieved by temporal modulation rather than frequency modulation of a single-photon comb with the same parameters. 

\subsubsection{Comparison to logical TFGKP qudit Bell states}\label{sec:gkp_bell_comp}
We can also make another natural comparison: to the qudit generalization of textbook superdense coding, using entangled TFGKP states that are logical Bell states of independently encoded TFGKP qudits. In the idealized case, a protocol of this kind would start with an initial state of the form
\begin{equation}
    \frac{1}{\sqrt{d}}(\ket{\overline{0}}\ket{\overline{0}}+\dots + \ket{\overline{d-1}}\ket{\overline{d-1}})
\end{equation}
for logical qudit dimension $d$. Encoding is performed by applying powers of the logical Pauli operators (the finite Heisenberg-Weyl operators $X$ and $Z$ defined in \eq{logical_X} and \eq{logical_Z}): 
\begin{equation}
    X^kZ^j \left[\frac{1}{\sqrt{d}}(\ket{\overline{0}}\ket{\overline{0}}+\dots + \ket{\overline{d-1}}\ket{\overline{d-1}})\right]
\end{equation}
For each unique pair of exponents $(j,k)$ of the Pauli operators, for $j, k \in \{0,\dots,d-1\}$, the initial state is mapped to one of the $d^2$ logical Bell states (all of which are stabilized by $X^d$ and $Z^d$). This encodes $2\log(d)$ bits of classical information. Thus we see that the capacity of this idealized scheme is twice the dimension $d$ of a single TFGKP qudit --- in other words, the square of the maximum number of distinguishable messages that can be encoded in a single-photon comb. However, the maximum number of distinguishable messages that can be encoded in a single-photon comb, for realistic experimental parameters, is precisely what was discussed in \sect{single_comb_comp} above. So we can use those numbers for the current comparison, as follows. 

Recall from \sect{single_photon} that the TFGKP qudit dimension is effectively the number of distinguishable states attainable by discrete frequency or temporal shifts of a single-photon comb (these states then define the logical $Z$ basis). Therefore one can choose either frequency or time to define the $Z$ basis, and it is advantageous to choose whichever one offers a larger number of distinguishable states, to maximize the resulting qudit dimension. 

We saw in \sect{single_comb_comp} that for the same experimental parameters that we assumed for our protocol, the largest number of distinguishable TFGKP states in a single-photon comb is $2^{1.93}$ distinguishable states, achievable by temporal shifts with a layer of classical error correction \footnote{To avoid confusion, note that the classical error correction layer here precedes the TFGKP code, and is used to \textit{define} the distinguishable states which form the TFGKP codewords. This would not be necessary for idealized combs, but as discussed extensively in \sect{single_comb_comp}, experimental imprecision means that naively defined TFGKP codewords would have nontrivial overlap and not be perfectly distinguishable. Ultimately, we are using this model for an apples-to-apples comparison to our superdense coding protocol, not for quantum error correction.}. This corresponds to a capacity of $2 \times 1.93 = 3.86$ bits per transmitted photon for the generalized textbook superdense coding scheme outlined above (which is similar to the 4-bit capacity of the Kwiat-Weinfurter Bell-state discrimination scheme \citep{Kwiat1998,Wei2007,Wang2019}). Our protocol's capacity of 8.91 bits improves upon this by a factor of $2.25$.

\subsubsection{Comparison to prior superdense coding schemes}\label{sec:prior_sdc_comp}
In Table~\ref{tab:comparison}, we provide a comparison of channel capacities for superdense coding protocols calculated in this work and prior works \citep{Kwiat1998,Wei2007,Wang2019}. Here, we have achieved record channel capacity among recent superdense coding protocols, using biphoton frequency combs and our proposed scheme. In addition, here we provide a comparison of optical loss and maximum channel capacity of our proposed scheme versus the Kwiat-Weinfurter Bell-state measurement scheme in the context of superdense coding. For the Kwiat-Weinfurter scheme, the maximum channel limit is 4 bits per transmitted photon, with a typical optical loss of 6 dB based on linear optics of Bell-state measurement \citep{Barreiro2008,Williams2017,Hu2018,Calsamiglia2001,Kwiat1998,Wei2007,Wang2019}. In contrast, in our proposed FBS scheme, the maximum channel capacity is 8.91 bits per transmitted photon, while the optical loss of the FBS is around 5 dB \citep{Nussbaum2022}. Therefore, our proposed FBS scheme has a 2.2 times higher channel rate compared to the previously highest achieved channel capacity of the Kwiat-Weinfurter scheme \citep{Kwiat1998,Wei2007,Wang2019}, with a comparable optical loss. Note that for this loss comparison, we do not include other sources of loss like fiber coupling losses and detector efficiency for both schemes.

\begin{table}
    \centering
    \includegraphics[width=\textwidth]{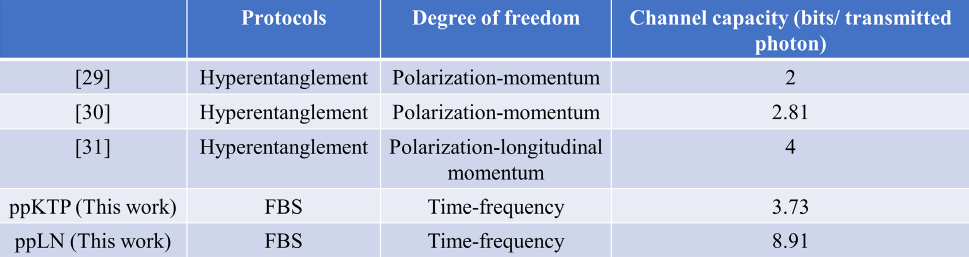}
    \caption{Summary of comparison of channel capacity (transmission rate) for superdense coding protocols.}
    \label{tab:comparison}
\end{table}

\section{Conclusion}\label{sec:conclusion}
We proposed a superdense coding protocol using entangled TFGKP states in the form of biphoton frequency combs, standard telecommunication components, time-resolving single-photon detectors, and a frequency beamsplitter, demonstrating the possibility of energy-time-entanglement-based superdense coding with currently achievable technologies. We estimated our scheme to be capable of sending 8.91 bits per photon, equivalent to 481 distinguishable messages, under realistic experimental conditions. Hence, our GKP-based transmission rate is 2.2 times higher than the highest previously achieved channel capacity of the Kwiat-Weinfurter scheme \citep{Kwiat1998,Wei2007,Wang2019} (more than an order-of-magnitude improvement in the number of distinguishable messages), with a comparable optical loss. Moreover, our results beat the rate achievable using a single-photon frequency comb with identical parameters by 4.6 times. This scheme encodes information into conjugate time and frequency degrees of freedom, and requires a frequency beamsplitter for decoding. Our approach has strong error resilience and operational simplicity using mostly passive devices. Consequently, this is a practical solution, particularly suitable for combining high-rate superdense coding with quantum communication tasks involving multidimensional time-frequency grid states.

\vspace{1em}

\noindent
\textbf{Acknowledgements}

\noindent
The authors acknowledge discussions with Yujie Chen, Hsiao-Hsuan Chin, and discussions on the superconducting nanowire single-photon detectors with Boris Korzh, Vikas Anant. This study is supported by the Army Research Office Multidisciplinary University Research Initiative (W911NF-21-2-0214), National Science Foundation under award numbers 1741707 (EFRI ACQUIRE), 1919355, 1936375 (QII-TAQS), and 2137984 (QuIC-TAQS). A.M. also acknowledges support from the Herb and Jane Dwight Stanford Graduate Fellowship.

\bibliographystyle{unsrtnat}
\bibliography{references}

\appendix

\section{Biphoton frequency comb in time basis}\label{appendix:app_a}
In this section we express the biphoton frequency comb state $\ketbfc$ from \eq{bfc_dirac} in the time basis via the Fourier transform, using the notation and time-frequency relations introduced in \sect{tf_review}. This makes manifest the asymmetry between the functional forms between the frequency and time wavefunctions of $\ketbfc$, reflecting the fact that $\ketbfc$ is not an eigenstate of any single-photon frequency shift $\fdisp{s}$ with $s\neq0$, while being an eigenstate of temporal shifts $\tdisp{s}$ when $s = 0$ mod $2\pi/\Delta\Omega$. All sums and integrals run from $-\infty$ to $\infty$.

\begin{align}
    \ket{\varphi_{00}} &= \sum_m \int d\Omega \,
    \delta(\Omega - m\Delta\Omega) \hat{a}_{H}^{\dagger}\left(\frac{\omega_p}{2} + \Omega\right) \hat{a}_{V}^{\dagger}\left(\frac{\omega_p}{2} - \Omega\right) \ket{\mathrm{v.s.}}
\\[4pt]
&= \sum_m \int d\Omega\, \delta(\Omega - m\DOmega)
  \frac{1}{2\pi}
  \int dt\, e^{-i(\frac{\omega_p}{2}+\Omega)t}
  a_H^\dagger(t)
  \int dt' e^{-i(\frac{\omega_p}{2}-\Omega)t'}
  a_V^\dagger(t') \ket{\mathrm{v.s.}}
\\[4pt]
&= \frac{1}{2\pi} \sum_m
    \int dt \int dt' \int d\Omega\,
   \delta(\Omega - m\DOmega)
   e^{-i\frac{\omega_p}{2}(t+t')}
   e^{-i\Omega(t-t')}
   a_H^\dagger(t) a_V^\dagger(t') \ket{\mathrm{v.s.}}
\\[4pt]
&= \frac{1}{2\pi} 
   \int dt \int dt'\,
   e^{-i\frac{\omega_p}{2}(t+t')} \sum_m e^{-i m \DOmega (t-t')}
   a_H^\dagger(t) a_V^\dagger(t') \ket{\mathrm{v.s.}}
\end{align}
Defining $\Delta T = \frac{2\pi}{\DOmega}$ and using the Poisson summation formula gives the expression
\begin{align}
&\, \int dt \int dt'\,
   e^{-i\frac{\omega_p}{2}(t+t')}
   \sum_m
   \delta\!\left((t-t') - m\Delta T\right)
   a_H^\dagger(t) a_V^\dagger(t') \ket{\mathrm{v.s.}}
\\[4pt]
&= \sum_m \int dt\,
   e^{-i\frac{\omega_p}{2}(2t-m\Delta T)}
   a_H^\dagger(t)
   a_V^\dagger(t-m\Delta T) \ket{\mathrm{v.s.}}
\\[4pt]
&= \sum_m \int dt \int d\tau\,
  e^{-i\frac{\omega_p}{2}(2t-\tau)}
   \delta(\tau-m\Delta T)
   a_H^\dagger(t)
   a_V^\dagger(t-\tau) \ket{\mathrm{v.s.}}.
\end{align}

\section{Alternative experimental configurations}\label{appendix:app_b}

\subsection{Alternative biphoton frequency comb state (singly-filtered configuration)}\label{sec:sfc_bfc}

\begin{figure}
  \centering
  \includegraphics[width=\linewidth]{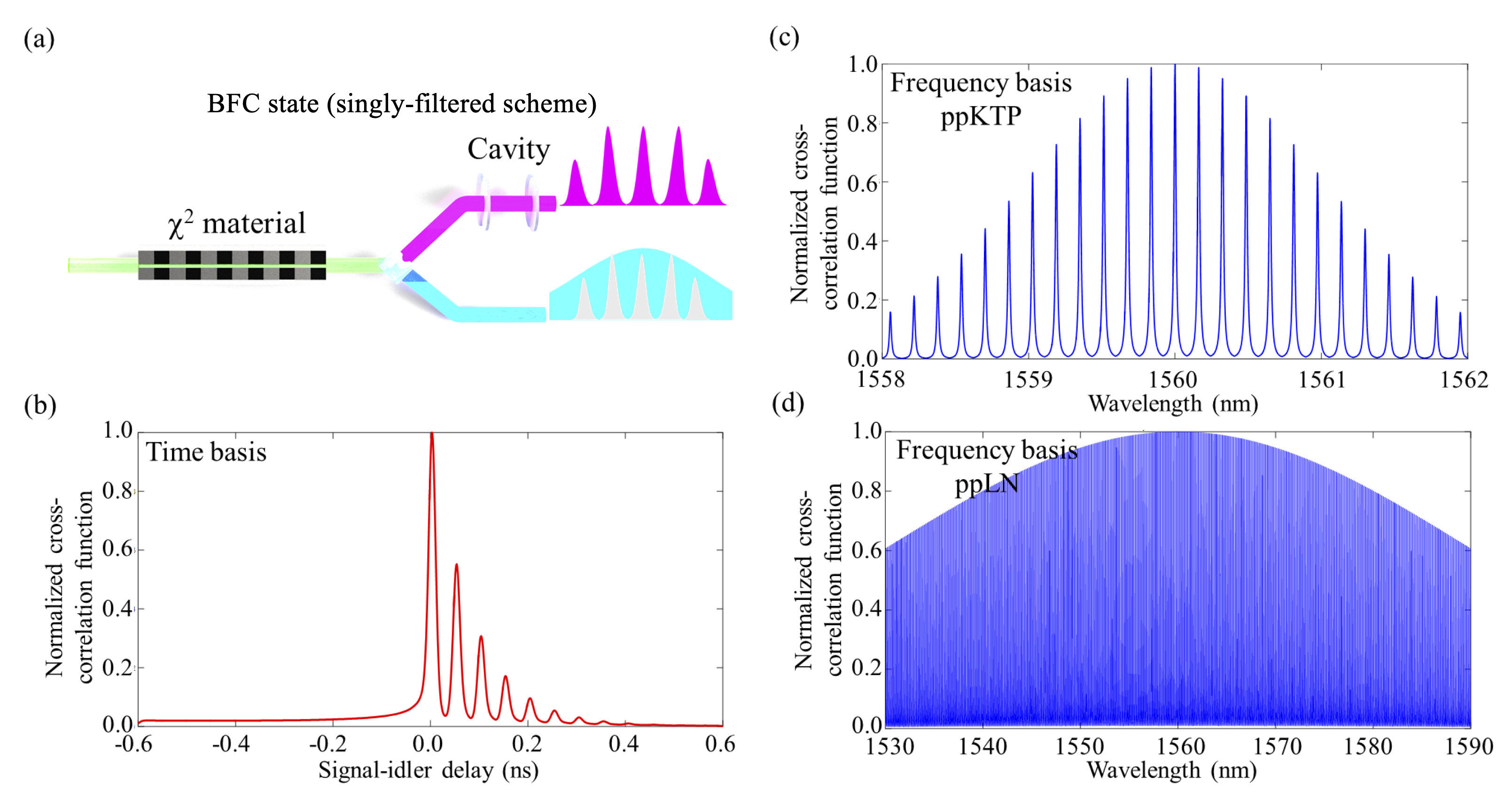}
  \caption{Modeling the probability distributions of a singly-filtered BFC in both the time and frequency bases. (a) Schematic of singly-filtered BFC state generation. (b) The signal-idler temporal second-order cross-correlation function versus relative time delay in BFC state with a 20~GHz FSR cavity. (c) Frequency spectrum of a BFC state, with phase-matching FWHM bandwidth of 250~GHz. (d) Frequency spectrum of a BFC state, with phase-matching FWHM bandwidth of 7.4~THz.}
  \label{fig:sfc_bfc}
\end{figure}

Here we provide a different biphoton frequency comb (BFC) state using a singly-filtered configuration \citep{Chang2023,Chang2021CLEO,Chang2025a,Chang2024b}. Fig.~\ref{fig:sfc_bfc} is an example of physical singly-filtered BFC states. Fig.~\ref{fig:sfc_bfc}(a) is the generation of BFC state via cavity-assisted scheme \citep{Chang2023,Chang2021CLEO,Chang2025a,Chang2024b}. We first calculate the time- and frequency-basis of a BFC state in a periodically-poled KTiOPO$_4$ (ppKTP) waveguide, as in Fig.~\ref{fig:sfc_bfc}(b) and \ref{fig:sfc_bfc}(c). Fig.~\ref{fig:sfc_bfc}(b) shows time-basis versus relative time delay in BFC state with a 20~GHz FSR cavity, a 2~GHz linewidth, and an FWHM timing jitter of 20~ps. The jitter is chosen to reveal the clear temporal discretized peaks of a BFC state, and this temporal oscillation of multiple peaks can support the phase-coherence among different frequency-bins in same state \citep{Chang2023,Chang2025a,Chang2024b,Ikuta2019}. In Fig.~\ref{fig:sfc_bfc}(c), we plot the frequency-basis of same BFC state in Fig.~\ref{fig:sfc_bfc}(b), with phase-matching FWHM bandwidth of 250~GHz. We can see that in the frequency-basis, our BFC state comprises of Gaussian and Lorentzian lineshapes. Moreover, in Fig.~\ref{fig:sfc_bfc}(d), we illustrate the BFC state using a broadband periodically poled lithium niobate (ppLN) with a FWHM bandwidth of 7.4~THz.

In the next section, we show how the experimental configuration for superdense coding can be modified to use a singly-filtered configuration, where only signal or idler photons are carved by a fiber cavity; in contrast to the main text, in which the BFC consists of a doubly-filtered BFC, where both biphotons are filtered. The detailed frequency-basis and temporal-basis wavefunction can be found in \citep{Chang2023,Chang2025a}. The singly-filtered BFC configuration causes the time-basis of a BFC state to have almost half the dimensions of a double-filtered BFC, although its frequency-basis has similar dimensions as doubly-filtered BFC. For all the calculations in Fig.~\ref{fig:sfc_bfc}, the center wavelength of the biphoton source is set to be 1560~nm.

\subsection{Proposed experimental configuration of singly-filtered BFC state for superdense coding protocol with FBS operation}\label{sec:sfc_sdc}

\begin{figure}
  \centering
  \includegraphics[width=\linewidth]{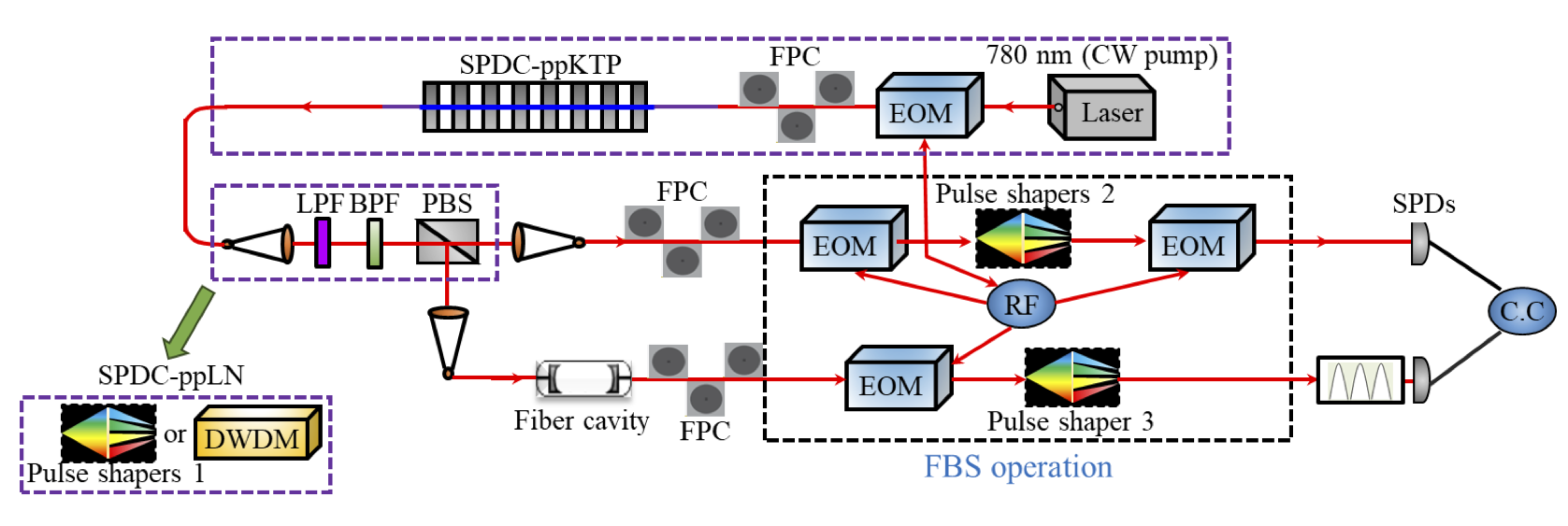}
  \caption{Proposed experimental setup for implementation of superdense coding with singly-filtered BFC and frequency beamsplitter (FBS) operation. Schematic of experimental configuration. CW, continuous-wave; EOM, electro-optic modulator; LPF, long-pass filter; BPF, band-pass filter; PBS, polarization beamsplitter; DWDM, dense wavelength-division-multiplexing. Here we designed our superdense coding scheme for both ppKTP and ppLN sources. For a type-II ppKTP, we can use a combination of LPF, BPF and PBS to reject pump photons while separating signal and idler photons. For a type-0 ppLN, we can use either pulse shaper 1 or DWDM to deterministically split biphotons in terms of their frequencies. Similar to \fig{experimental_setup} of the main text, all the EOMs are designed to synchronize with a single RF source. An additional tunable frequency filter is designed to carry out time and frequency measurement for superdense encoding via two single-photon detectors. The purple highlighted regions are the standard measurement setup for generating a BFC state.}
  \label{fig:sfc_sdc}
\end{figure}

In this section, we propose and design the experimental implementation of superdense coding with singly-filtered BFC states and a near deterministic frequency beamsplitter (FBS) operation \citep{Lu2018,Seshadri2022}. Fig.~\ref{fig:sfc_sdc} is the proposed experimental setup. The design consideration is different from that of doubly-filtered BFC in the main text. Here we designed our superdense coding scheme for both ppKTP and ppLN sources. For a type-II ppKTP, we can use a combination of long-pass filter (LPF), band-pass filter (BPF) and polarization beamsplitter (PBS) to reject pump photons while separating signal and idler photons. For a type-0 ppLN, we can use either pulse shaper 1 or dense wavelength-division-multiplexing (DWDM) to deterministically split biphotons in terms of their frequencies. After the separation of signal and idler photons, we connect a fiber cavity on idler photons to generate BFC state via a singly-filtered BFC configuration. Then, the first electro-optic modulator (EOM) is for providing frequency shifts (and temporal shifts) for pump wavelength, when combined with FBS operation, to distinguish complete frequency-bin states \citep{Seshadri2022}. Near-deterministic frequency beamsplitter (FBS) operation can be achieved with a sandwiched EOM-pulse shaper-EOM scheme \citep{Lu2018,Seshadri2022}. Similar to \fig{experimental_setup} of the main text, all the EOMs are designed to synchronize with a single RF source. An additional tunable frequency filter is designed to carry out frequency measurement for superdense encoding protocol. The purple highlighted regions are the standard measurement setup for generating singly-filtered BFC states. The advantage of this scheme is that there is smaller filtering loss from a cavity (only idler is carved here, for example), however, in order to realize superdense coding with FBS operation, it is necessary to use more EOMs and pulse shapers (incur more insertion losses) for both signal and idler photons after their separation.

\subsection{Proposed experimental configuration for superdense coding with frequency-polarization hyperentanglement using BFC states}\label{sec:fp_sdc}

\begin{figure}
  \centering
  \includegraphics[width=\linewidth]{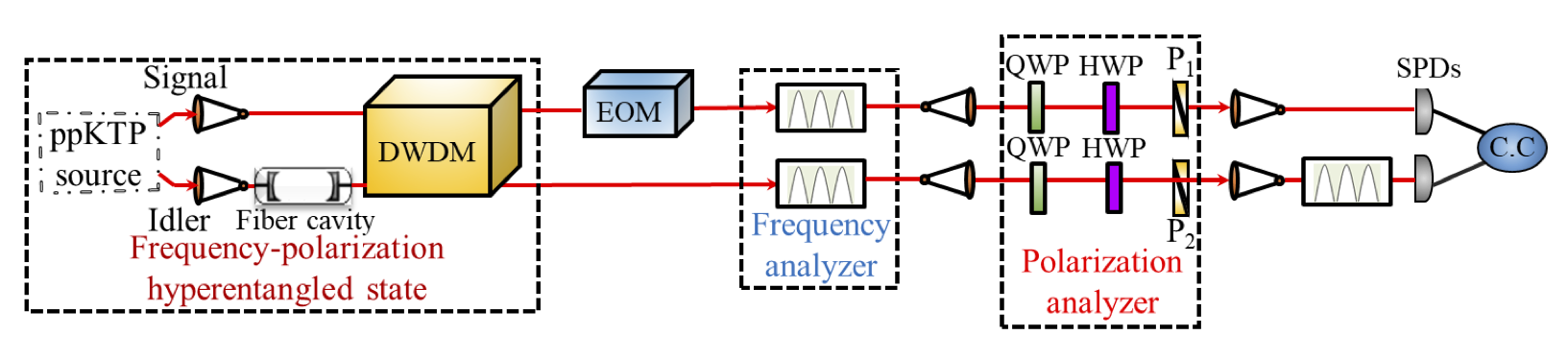}
  \caption{Proposed experimental setup for implementation of superdense coding with singly-filtered BFC and frequency-polarization hyperentanglement. QWP, quarter-wave plate; HWP, half-wave plate; P, polarizer. We propose to generate frequency-polarization hyperentangled BFC states with a singly-filtered BFC configuration. The frequency shift is enabled by an EOM. After the encoding process, we use the cascaded frequency and polarization analyzers to decode the shifted BFC states. An additional tunable frequency filter is designed to carry out time and frequency measurement to conclude the superdense encoding via two SPDs.}
  \label{fig:fp_sfc_sdc}
\end{figure}

Alternatively, we can also generate frequency-polarization hyperentangled state from a singly-filtered BFC. In Fig.~\ref{fig:fp_sfc_sdc}, first, after a type-II ppKTP source, we connect idler photons to a cavity to generate singly-filtered BFC (BFC states in Fig.~\ref{fig:sfc_bfc}). Then, similar to Fig.~\ref{fig:sfc_sdc}, we use the EOM to induce the required frequency shift for encoding process of superdense coding. Then, the rest of the decoding process is the same as in Fig.~\ref{fig:fp_sfc_sdc}. The main difference is here, the cavity filtering only occurs to half of biphotons, so this scheme has higher flux for faster data acquisition. For the channel capacity of the singly-filtered BFC, one could potentially exploit the structure of the time-domain wavefunction to increase the capacity compared to the doubly-filtered case. At least, we do not expect the capacity to decrease for the same parameters, since the number of distinguishable temporal shifts is determined by the periodicity of the quantum comb.

After having introduced the singly-filtered BFC for superdense coding protocols, we now turn our focus back to the doubly-filtered BFC case. As can be seen in the main text \eq{bfc_dirac} and \eq{bfc_exp_freq}, the doubly-filtered BFC state we introduced can naturally carry frequency-polarization hyperentanglement, which has been experimentally confirmed in prior works \citep{Xie2015,Chang2021}. This hyperentanglement can be used to implement superdense coding with a pair of ancilla polarization qubits \citep{Barreiro2008,Williams2017,Hu2018}. Previous demonstrations have showcased implementations of Bell-state measurements using hyperentangled states in both orbital angular momentum and polarization \citep{Barreiro2008}, time and polarization \citep{Williams2017}, as well as states hyperentangled in path and polarization \citep{Hu2018}. However, these prior works do not utilize the discretized frequency DoF, which is advantageous for efficient transmission and multiplexing through fiber-based quantum networks \citep{Wengerowsky2018,Joshi2020}. Here we propose and design the usage of the frequency-polarization hyperentanglement for the encoding process of superdense coding. The hyperentanglement protocol has two independent components. The time-frequency encoding is the same as for the superdense coding protocol but with no temporal shifts permitted, eliminating the need for an FBS. In addition, the BFC state preparation is slightly modified such that the polarization DoF of the two photons are maximally entangled and can be used to implement the standard 2 bits per transmitted qubit superdense coding protocol.

\begin{figure}
  \centering
  \includegraphics[width=\linewidth]{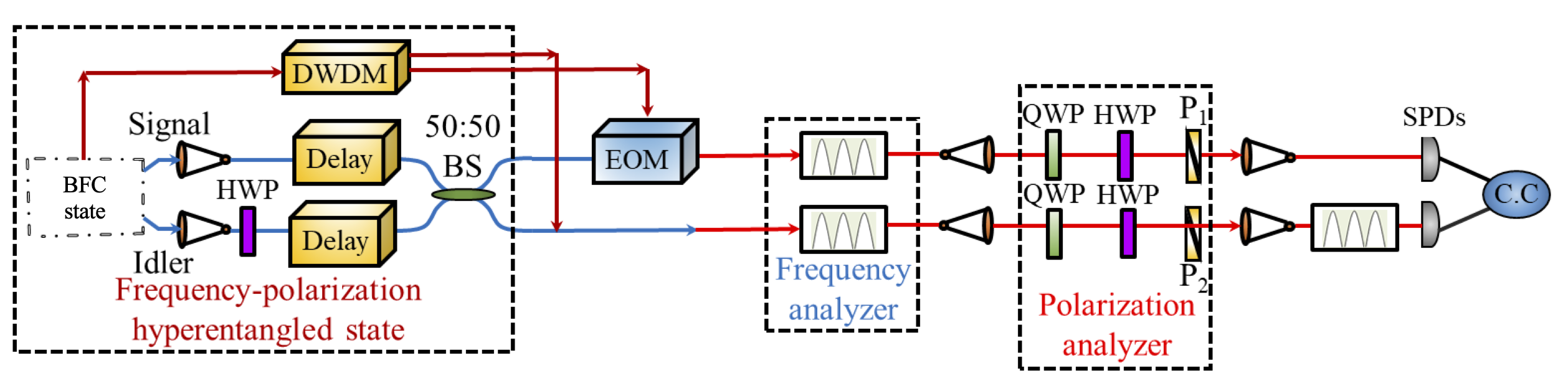}
  \caption{Experimentally designed setups for implementation of superdense coding with BFC states and hyperentanglement. The second method is based on frequency-polarization hyperentanglement. EOM, electro-optic modulator; TDCM, tunable dispersion compensation module; SPDs, single-photon detectors; C.C., coincidence counts. The frequency shift is enabled by an EOM. After the encoding process, we use the cascaded frequency and polarization analyzers to distinguish the hyperentangled states. DWDM, dense wavelength-division-multiplexing. QWP, quarter-wave plate; HWP, half-wave plate; P, polarizer. For both schemes, additional tunable frequency filter is designed carry out time and frequency measurement to conclude the superdense coding protocols via two SPDs.}
  \label{fig:fp_dfc_sdc}
\end{figure}

In Fig.~\ref{fig:fp_dfc_sdc}, we lay out the experimental considerations of a BFC state with frequency-polarization hyperentanglement for superdense coding. Initially, our BFC state can be generated using a cavity-filtered nonlinear optical process \citep{Xie2015,Chang2021}. Here we use a fiber cavity to carve the frequency spectrum of a SPDC using a ppKTP waveguide. Due to a type-II configuration of our source, we can generate frequency-polarization hyperentangled states via a post-selection method, where we mix orthogonal polarization photons in a 50:50 BS \citep{Xie2015,Chang2021}. We employ two adjustable fiber delay lines for the purpose of aligning the temporal wavepacket of signal and idler photons. Then, we can incorporate a half-wave plate (HWP) preceding the 50:50 BS to change the polarization of idler photons to generate frequency-polarization hyperentanglement. Alternatively, in Fig.~\ref{fig:fp_dfc_sdc}, we also propose to generate deterministic frequency-polarization hyperentanglement via a dense wavelength-division-multiplexing (DWDM) \citep{Kaiser2012}. If using a type-0 ppLN biphoton source, we can use Sagnac interferometry to generate frequency-polarization hyperentanglement \citep{Wengerowsky2018,Joshi2020}. Owing to the discretization of the BFC states, we expect Hong-Ou-Mandel (HOM) recurrence interferences after sending our BFC states into a 50:50 BS \citep{Xie2015,Chang2021}. Also, in the same setup, it is possible to use frequency-filtered HOM interference (or spatial quantum beating \citep{Zhang2021}) to encode different frequency bin states. As in \fig{experimental_setup} of the main text, an EOM enables the frequency shift. After this shift, we direct biphotons into the cascaded frequency and polarization analyzers to decode the transmitted information. We also insert an extra tunable frequency filter before the detectors to recover the frequency shift in the encoding process. 

The analysis of the transmission capacity for frequency portion of this scheme is similar to that of the frequency decoding of the FBS protocol in the main text, leading to a transmission rate of 6.98 bits per transmitted photon at vanishing error, achieved for $c$ of 202. Neglecting errors, the maximum transmission rate for the polarization DoF will be 2 bits per transmitted photon. Therefore, in total, the scheme will achieve 8.98 bits per transmitted photon. This is comparable to the FBS scheme, thanks to the capacity boost provided by the additional polarization entanglement.

\end{document}